\documentclass[12pt]{article}
\usepackage{amsmath}
\usepackage{graphicx,psfrag,epsf}
\usepackage{enumerate}
\usepackage{natbib}
\usepackage{url} 

\usepackage{subcaption}

\newcommand{\blind}{1}

\usepackage{float}
\usepackage{times}
\usepackage{bm}
\usepackage{amssymb}
\usepackage{mathrsfs}
\usepackage{amsthm}
\usepackage{hyperref}
\usepackage{cancel}
\usepackage{mathtools}
\usepackage{enumitem}

\newtheorem{theorem}{Theorem}
\newtheorem{definition}{Definition}
\newtheorem{lemma}{Lemma}

\begin{document}

\def\spacingset#1{\renewcommand{\baselinestretch}%
{#1}\small\normalsize} \spacingset{1}


\if1\blind
{
  \title{\bf The Cox-Polya-Gamma Algorithm for Flexible Bayesian Inference of Multilevel Survival Models}
  \author{Benny Ren\\
    Department of Family, Population and Preventive Medicine \\
    Renaissance School of Medicine at Stony Brook University, Stony Brook, NY\\
     \\
    Jeffrey S. Morris and Ian Barnett \\
    Department of Biostatistics, Epidemiology, and Informatics\\
    Perelman School of Medicine at the University of Pennsylvania, Philadelphia, PA}
  \maketitle
} \fi

\if0\blind
{
  \bigskip
  \bigskip
  \bigskip
  \begin{center}
    {\LARGE\bf The Cox-Polya-Gamma Algorithm for Flexible Bayesian Inference of Multilevel Survival Models}
\end{center}
  \medskip
} \fi

\bigskip
\begin{abstract}
Bayesian Cox semiparametric regression is an important problem in many clinical settings. Bayesian procedures provide finite-sample inference and naturally incorporate prior information if MCMC algorithms and posteriors are well behaved. Survival analysis should also be able to incorporate multilevel modeling such as case weights, frailties and smoothing splines, in a straightforward manner. To tackle these modeling challenges, we propose the Cox-Polya-Gamma (Cox-PG) algorithm for Bayesian multilevel Cox semiparametric regression and survival functions. Our novel computational procedure succinctly addresses the difficult problem of monotonicity constrained modeling of the nonparametric baseline cumulative hazard along with multilevel regression. We develop two key strategies. First, we exploit an approximation between Cox models and negative binomial processes through the Poisson process to reduce Bayesian computation to iterative Gaussian sampling. Next, we appeal to sufficient dimension reduction to address the difficult computation of nonparametric baseline cumulative hazard, allowing for the collapse of the Markov transition within the Gibbs sampler based on beta sufficient statistics. In addition, we explore conditions for uniform ergodicity of the Cox-PG algorithm. We demonstrate our multilevel modeling approach using open source data and simulations. We provide software for our Bayesian procedure in the supplement.
\end{abstract}
\vfill

\newpage
\spacingset{1.4} 

\section{Introduction}
Semiparametric Cox proportional hazards (PH) regression is an important tool used for time to event data \citep{cox1972regression,cox1975partial}. Many extensions of the Cox model have been proposed, such as cure models (or zero-inflation) \citep{sy2000estimation}, frailty models (or random effects) \citep{wienke2010frailty}, smoothing splines \citep{therneau2000cox,cui2021additive}, convex set constraints \citep{mcgregor2020cox}, variable selection \citep{tibshirani1997lasso},  and other methods. However, nonparametric inference for semiparametric Cox regression involves difficult monotonicity constrained inference of the cumulative hazard \citep{hothorn2018most}. The Bayesian paradigm provides a convenient framework to naturally address many of these variations of the canonical Cox model \citep{ibrahim2001bayesian,chen2006posterior}. Bayesian methods can naturally incorporate convex set constraints \citep{valeriano2023moments,pakman2014exact,maatouk2016new} that can be applied to estimation and inference of monotonic cumulative hazard functions. In addition, Bayesian hierarchical modeling can be extended to different types of regression, such as random effects \citep{wang2018analysis}, zero-inflation \citep{neelon2019bayesian}, smoothing splines \citep{wand2008semiparametric}, and variable selection \citep{carvalho2009handling,bhadra2019lasso,makalic2015simple} with straightforward Gibbs samplers for multilevel Gaussian models. In order to leverage this rich resource, we propose a novel Markov chain Monte Carlo (MCMC) algorithm that allows Cox model parameters to be modeled with hierarchical Gaussian models 

The Cox model can be described as locally a Poisson process that is constrained to binary outcomes of recording death or right censoring \citep{ibrahim2001bayesian,aalen2008survival}. In recent years, \citealt{polson2013bayesian} proposed a Polya-gamma (PG) Gibbs sampler for logistic and negative binomial (NB) regression that is based in Gaussian sampling. Consequently, the negative binomial process is a useful analog to the Poisson process of Cox regression. We use a standard gamma process Bayesian modification to Cox regression that results in a NB kernel and enables Gaussian-based modeling to incorporate the necessary structure and constraints. By specifying a univariate gamma frailty or gamma prior on the cumulative hazard, we establish an analogous frailty model to locally negative binomial relationship \citep{aalen2008survival,winkelmann2008econometric}. Typically, a vague gamma frailty is specified to bridge Poisson and NB models \citep{kalbfleisch1978non,sinha2003bayesian,duan2018scaling}. The practicality of a vague gamma prior allows access to the PG sampler, benefiting from the use of Gaussian Gibbs samplers. In addition, maintaining Gibbs transitions is crucial as we can take advantage of the marginal transitions to derive succinct Metropolis-Hastings (MH) acceptance probabilities that can be be used to remove bias due to gamma frailty augmentation \citep{duan2018scaling}.

However, the NB form alleviates much of the algebra associated with the cumulative hazard but does not address the entirety of the baseline hazard rate which has positivity constraints. There are a broad class nonparametric estimators that rely on partitioning the time interval  \citep{nieto2002markov,nelson1972theory,aalen1978nonparametric,kaplan1958nonparametric,ibrahim2001bayesian}. Evaluating these estimators without considering the computational algorithm often results in difficult MCMC procedures due to monotonic constraints on the baseline cumulative hazard with MH steps that require tuning. There are many tools available to enforce convex set constraints on Gaussian sampling \citep{valeriano2023moments,pakman2014exact,maatouk2016new,cong2017fast,li2015efficient} which we will demonstrate can be used to enforce monotonicity and positivity constraints of Cox regression. Inspired by previous work in nonparametric analysis, we proposed the use of a sufficient statistic partition-based estimator with closed forms for Gibbs sampler conditional distributions. In order to do so, we introduce a novel monotonic spline system with a half-space constraint \citep{ramsay1988monotone,meyer2018convergence} to model the baseline log cumulative hazard function that factorizes into beta sufficient statistics in the likelihood. Next, we introduce slice sampler auxiliary variables to remove the sufficient statistics associated with the baseline hazard rate from the algebra \citep{damien1999gibbs,mira2002efficiency,neal2003slice}. Our spline system collapses the linear operator of our Markov transitions on to beta sufficient statistics to improve the efficiency of Bayesian computation while retaining the Gaussian structure. This results in a flexible Bayesian procedure we refer to as the Cox-Polya-Gamma (Cox-PG) algorithm that is appealing for being able to adapt multilevel Gaussian inference to Cox regression.

Aside from being able to incorporate prior information into the Cox model, our Cox-PG approach is desirable for situations where inference on the baseline hazard is necessary such as Kaplan-Meier models \citep{kaplan1958nonparametric}. We also show that case weights such as inverse probability weighting and power priors, can be incorporated into the Cox-PG algorithm through fractal beta kernels  \citep{shu2021variance,ibrahim2015power}.  In addition, we explore conditions under which the Cox-PG sampler is uniformly ergodic. Consequently, we are able to derive ergodic theory results for our Gibbs algorithm under mild assumptions. Hierarchical models from Gaussian Gibbs samplers are natural extensions of our approach with straightforward modifications to the Cox-PG algorithm. This includes cure models or zero-inflation \citep{sy2000estimation,neelon2019bayesian}, joint models \citep{wulfsohn1997joint,wang2022time}, copulas \citep{wienke2010frailty}, competing risk \citep{kalbfleisch2011statistical} and variable selection \citep{carvalho2009handling,bhadra2019lasso,makalic2015simple} and other multilevel Gaussian models. As a result, we propose a unified Bayesian framework that envelopes many of the models found throughout survival analysis \citealt{ibrahim2001bayesian}. In contrast to influence function based semiparametric theory (\citealt{tsiatis2006semiparametric} Chapter 5), the Cox-PG sampler provides a flexible inference procedure for finite sample Cox modeling, addressing a critical gap in many survival analysis settings such as clinical trials \citep{ildstad2001small}.

The rest of the article is organized as follows. In Section \ref{method}, we outline the Cox-PG algorithm. In Section \ref{theory}, we present uniform ergodicity theoretical results. We present additional methods for accelerating MCMC algorithms, removing bias, and variations of Cox semiparametric regression in Section \ref{accel} and \ref{misc}. Simulation studies and real data analysis involving multilevel survival models are presented in Section \ref{examples}, \ref{sim} and \ref{leuk}.

\section{Method: Cox-PG Gibbs sampler} \label{method}
Semiparametric Cox PH regression is closely related to the exponential family with an additional hazard rate term that needs to be accounted for with constrained inference. Exponential family Bayesian models are well studied and have appealing theoretical properties \citep{choi2013polya}. In order to emphasize Cox model constraints and its Poisson kernel, we show the derivation of the transformation model for Cox PH regression \citep{hothorn2018most}. The extreme value distribution function of PH models is given as $F_{\mathrm{MEV}}\{A(t)\}=1- \exp [ -\exp \{A(t)\} ]$ with survival function $S_{\mathrm{MEV}}\{A(t)\}=\exp [ -\exp \{ A(t)\} ]$. In this case, study time $T_i$ denotes the censoring or event time for subject $i \in \{1,\dots,N\}$. Let $A(t_i) = \log \{ \Lambda(t_i) \} + \mathbf{x}^\top_i \boldsymbol{\beta} = \alpha(t_i) + \mathbf{x}^\top_i \boldsymbol{\beta} = \mathbf{z}^{\top}_{\alpha}(t_i) \mathbf{u}_{\alpha} + \mathbf{x}^\top_i \boldsymbol{\beta} = \mathbf{m}_{i}^{\top} \boldsymbol{\eta}$. Monotonic function $\alpha(t_i) = \sum^J_{j=1} z_{\alpha,j}(t_i) u_{\alpha,j} =  \mathbf{z}^{\top}_{\alpha}(t_i) \mathbf{u}_{\alpha}$ is the baseline log cumulative hazard approximated using additive monotonic splines with the constraints $\mathbf{u}_{\alpha} > \mathbf{0}$ represented through $\mathcal{C}_0 = \left\{ \boldsymbol{\eta} :  \mathbf{u}_{\alpha} > \mathbf{0} \right\}$. The baseline hazard is $D_t \Lambda(t_i) =  D_t \exp \left\{ \alpha_0 + \mathbf{z}^{\top}_{\alpha}(t_i) \mathbf{u}_{\alpha} \right\} = \left\{ D_t \mathbf{z}^{\top}_{\alpha}(t_i) \mathbf{u}_{\alpha} \right\} \exp \left\{ \alpha_0 + \mathbf{z}^{\top}_{\alpha}(t_i) \mathbf{u}_{\alpha} \right\} $ where the differential operator is represented as $d/dt = D_t$ and $\alpha_0$ is an intercept incorporated into $\mathbf{x}^\top_i \boldsymbol{\beta}$. The marginal likelihood contribution of the model (\citealt{kalbfleisch2011statistical} Chapter 2) is 
$$
\begin{array}{c}
\left[ D_t F_{\mathrm{MEV}} \{ A(t_i) \} \right]^{y_i} \left[ S_{\mathrm{MEV}} \{ A(t_i) \} \right]^{1-y_i} = \left\{ D_t \alpha(t_i) \right\}^{y_i} \exp \{y_i A(t_i)\} S_{\mathrm{MEV}}\{A(t_i)\}
\end{array}
$$
with death (or event) variable $y_i$ and right censoring $1-y_i$. The hazard function is given by $\left\{ D_t \alpha(t_i) \right\}^{y_i} \exp \{y_i A(t_i)\}$ following the canonical factorization of hazard-survival functions. The flexibility of this framework is seen by noting that if $\alpha(t) = \alpha_0 + \log(t)$, the likelihood is an exponential PH regression while if $\alpha(t) = \alpha_0 + \gamma \log(t)$ we have the Weibull PH model and the parameterization of $\alpha(t)$ determines the hazard family. Now, we set $A(t_i) = \mathbf{m}^\top_i \boldsymbol{\eta}$, where $\boldsymbol{\eta} = \left[ \mathbf{u}^\top_{\alpha}, \boldsymbol{\beta}^\top \right]^\top$ and $\mathbf{m}_i = \left[ \mathbf{z}^\top_{\alpha}(t_i), \mathbf{x}^\top_i \right]^\top$, and the resulting generalized likelihood is 
$$
\begin{array}{c}
L_{\mathrm{PH}}(\boldsymbol{\eta} \mid \mathbf{y}, \mathbf{M} ) =  \left[ \prod_{i=1}^N \left\{ D_t \mathbf{z}_{\alpha}^{\top} (t_i) \mathbf{u}_\alpha \right\}^{y_i} \right] \mathbb{I}( \boldsymbol{\eta} \in \mathcal{C}_0 ) \exp \left( \mathbf{y}^{\top} \mathbf{M} \boldsymbol{\eta} \right) \exp \left\{ -\sum_{i=1}^{N} \exp \left(\mathbf{m}_{i}^{\top} \boldsymbol{\eta}\right) \right\} .
\end{array}
$$ 
The derivative of monotonic splines, $D_t \mathbf{z}^{\top}_{\alpha}(t_i)$, is positive and the basis coefficients are constrained to be positive. The design matrix $\mathbf{M} \in \mathbb{R}^{N \times (J+P)}$ is composed of rows $\mathbf{m}_{i}$. Under this construction, slice sampler auxiliary variables are suitable to maintain positive $D_t \mathbf{z}^{\top}_{\alpha}(t_i) \mathbf{u}_{\alpha} \in \mathbb{R}_+ \coloneqq (0,\infty)$ \citep{damien1999gibbs,mira2002efficiency}.

Slice samplers naturally remove positive terms in the likelihood by enforcing constraints in conditional distributions. In order to effectively incorporate a slice sampler, we start by placing a gamma mixture variable $z_{\gamma,i} \sim \mathrm{G}(\epsilon, \epsilon)$ on the Poisson kernel of PH models to obtain a frailty model analogous to negative binomial regression. Note that the first two moments are given as $\mathbb{E}(z_{\gamma,i})=1$ and $\mathrm{Var}(z_{\gamma,i})=\epsilon^{-1}$ and a vague mixture $\epsilon \gg 0$ can be used. As noted in \cite{duan2018scaling} this limit approximation works well for Poisson processes. In addition, we show how to remove this bias later on in the article through a simple MH step. A quick Laplace transform reveals the modification to the baseline hazard under frailty: $D_t \Lambda_{\text{frailty}}(t) = D_t \Lambda(t)/\left\{1 + \epsilon^{-1} \Lambda(t) \right\}$ (\citealt{aalen2008survival} Chapter 6) and by sending $\epsilon \rightarrow \infty$ we have the Cox model hazard. We remove the hazard component of the likelihood with multiplication of slice sampling auxiliary uniform variable $\nu_i \mid \boldsymbol{\eta} \sim \mathrm{U} \{0, D_t \mathbf{z}^{\top}_{\alpha}(t_i) \mathbf{u}_{\alpha} \}$ for uncensored events, $y_i=1$. The resulting likelihood is given as 
$$
\begin{array}{rl}
L_{\text{PH}}(\boldsymbol{\eta}, \mathbf{z}_{\gamma} \mid \mathbf{y}, \mathbf{M} ) \pi(\boldsymbol{\nu} \mid \boldsymbol{\eta} ) \pi( \mathbf{z}_{\gamma} ) =& 
\left[ \prod_{i=1}^N \mathbb{I} \left\{ \nu_i \leq D_t \mathbf{z}^{\top}_{\alpha}(t_i) \mathbf{u}_{\alpha} \right\}^{y_i} \right]
\mathbb{I}( \boldsymbol{\eta} \in \mathcal{C}_0 ) \\
& \times \left[ \prod_{i=1}^N \left\{ \exp \left( \mathbf{m}_{i}^{\top} \boldsymbol{\eta} \right) z_{\gamma,i} \right\}^{y_i} \right] \pi( \mathbf{z}_{\gamma} ) \\
&\times \exp \left\{ -\sum_{i=1}^{N} \exp \left(\mathbf{m}_{i}^{\top} \boldsymbol{\eta}\right) z_{\gamma,i} \right\}
\end{array}
$$
where $\mathbf{z}_\gamma$ is a vector of the $z_{\gamma,i}$ (\citealt{wienke2010frailty} Chapter 3) and $\nu_i \geq 0$. We can marginalize out $\mathbf{z}_\gamma$ by carrying out the Mellin transform to obtain a negative binomial kernel
$$
\begin{array}{c}
\frac{1}{y_i !} \int_0^{\infty} \exp(-\lambda_i {z_\gamma} ) \lambda_i^{y_i} z_\gamma^{y_i} \frac{\epsilon^\epsilon}{\Gamma(\epsilon)} z_\gamma^{\epsilon-1} e^{-\epsilon z_\gamma} d z_\gamma 
= \frac{ \Gamma(y_i+\epsilon) }{ \Gamma(y_i+1) \Gamma(\epsilon) }\left(\frac{ \epsilon }{\epsilon + \lambda_i}\right)^{\epsilon} \left(\frac{ \lambda_i }{\epsilon + \lambda_i}\right)^{y_i}.
\end{array}
$$
Let $\psi_i = \mathbf{m}_{i}^{\top} \boldsymbol{\eta} - \log(\epsilon) = \mathbf{m}_{i}^{\top} \boldsymbol{\eta} + \eta_\epsilon$ and $\lambda_i = \exp( \mathbf{m}_{i}^{\top} \boldsymbol{\eta} )$, then the NB likelihood is
\begin{equation}\label{NBlik}
\begin{array}{rl}
L_{\text{NB}}(\boldsymbol{\eta} \mid \mathbf{y}, \mathbf{M} ) \propto& \left[ \prod_{i=1}^N \left\{ D_t \mathbf{z}^{\top}_{\alpha}(t_i) \mathbf{u}_{\alpha} \right\}^{y_i} \right]
\mathbb{I}( \boldsymbol{\eta} \in \mathcal{C}_0 ) \left[ \prod_{i=1}^N (p_i)^{y_i} (1 - p_i)^{\epsilon} \right]\\
L_{\text{NB}}(\boldsymbol{\eta} \mid \mathbf{y}, \mathbf{M} ) \pi(\boldsymbol{\nu} \mid \boldsymbol{\eta} ) \propto& 
\left[ \prod_{i=1}^N \mathbb{I} \left\{ \nu_i \leq D_t \mathbf{z}^{\top}_{\alpha}(t_i) \mathbf{u}_{\alpha} \right\}^{y_i} \right]
\mathbb{I}( \boldsymbol{\eta} \in \mathcal{C}_0 ) \left[ \prod_{i=1}^N (p_i)^{y_i} (1 - p_i)^{\epsilon} \right]
\end{array}
\end{equation}
where $p_i = \mathrm{expit}( \psi_i ) = \exp(\psi_i)/\{ 1 + \exp(\psi_i) \}$ and we condition on $\epsilon$. We can integrate out $\nu_i \geq 0$ to obtain the proportional frailty likelihood. The frailty likelihood is multiplied by constant $\prod_{i=1}^N \epsilon^{-y_i}$ to get the NB likelihood. After accounting for covariates and conditioning on $\epsilon$, through the Laplace transform, the hazard function is proportional to $D_t \Lambda_{\text{frailty}}(t_i \mid -) \propto \left\{ D_t \mathbf{z}^{\top}_{\alpha}(t_i) \mathbf{u}_{\alpha} \right\}^{y_i} p_i^{y_i}$, which allows for data augmentation (\citealt{wienke2010frailty} Chapter 3). The survival function under frailty is $(1-p_i)^\epsilon$ resulting in the canonical hazard-survival factorization. 

To account for the logit link, we can use a PG auxiliary variable $\omega \mid b,c \sim \mathrm{PG}\left(b, c\right)$, where $b>0$ \citep{polson2013bayesian}. The PG density is 
$$
\begin{array}{c}
\pi(\omega \mid b,c) = \frac{1}{2 \pi^2} \sum_{k=1}^{\infty} \frac{g_k}{(k-1 / 2)^2+c^2 /\left(4 \pi^2\right)}
\end{array}
$$
where $g_k$ are independently distributed according to $\mathrm{G}(b, 1)$. We have the following exponential tilting or Esscher transform \citep{esscher1932probability} relationship for PG distributions
\begin{equation}\label{id1}
\begin{array}{c}
\frac{\left(e^\psi\right)^a}{\left(1+e^\psi\right)^b} = 2^{-b} e^{\kappa \psi} \int_0^{\infty} e^{-\omega \psi^2 / 2} \pi (\omega \mid b, 0) d \omega, \quad
\pi (\omega \mid b, c)=\frac{\exp \left(-c^2 \omega / 2\right) \pi (\omega \mid b, 0)}{\mathbb{E}_{ \omega \mid b,0 }\left[\exp \left(-c^2 \omega / 2\right)\right]}
\end{array}
\end{equation}
where $\kappa=a-b / 2$ and $\pi(\omega \mid b, 0)$ denotes a $\mathrm{PG}(b, 0)$ density. We use $\omega_i \sim \mathrm{PG}\left(y_i+\epsilon, 0 \right)$ and get
$$
\begin{array}{rl}
\frac{\left(e^{\psi_i} \right)^{y_i}}{\left(1+e^{\psi_i} \right)^{y_i+\epsilon}} \propto \int_0^{\infty}
e^{\kappa_i \psi_{i}} e^{-\omega_{i} \psi_{i}^2 / 2} \pi \left(\omega_{i} \mid y_i + \epsilon, 0\right) d \omega_{i} 
\end{array}
$$
to tease out a Gaussian kernel. Note that 
$$
\begin{array}{c}
\pi(\omega_i \mid y_i + \epsilon, \psi_i)=\cosh^{y_i+\epsilon} ( |\psi_i / 2| ) \exp \left( - \omega_i \psi_i^2 /2 \right) \pi (\omega_i \mid y_i+\epsilon, 0),
\end{array}
$$
can be shown with a Laplace transform \citep{polson2013bayesian,choi2013polya}.
We get the NB posterior
\begin{equation}\label{lik1}
\begin{array}{rl}
L_{\text{NB}}(\boldsymbol{\eta} \mid \mathbf{y}, \mathbf{M} ) \pi(\boldsymbol{\nu}, \boldsymbol{\omega} \mid \boldsymbol{\eta} ) \propto& 
\left[ \prod_{i=1}^N \mathbb{I} \left\{ \nu_i \leq D_t \mathbf{z}^{\top}_{\alpha}(t_i) \mathbf{u}_{\alpha} \right\}^{y_i} \right]
\mathbb{I}( \boldsymbol{\eta} \in \mathcal{C}_0 )  \\
& \times \prod_{i=1}^N e^{\kappa_i \psi_{i}} e^{-\omega_{i} \psi_{i}^2 / 2} \pi \left(\omega_{i} \mid y_i + \epsilon, 0\right)
\end{array}
\end{equation}
where $\kappa_i = (y_i-\epsilon)/2$. Integrating out $\boldsymbol{\nu}$ and $ \boldsymbol{\omega}$ results in the original NB likelihood. We can write our Gaussian density conditioned on $\eta_\epsilon$ through $
\begin{array}{c}
e^{\kappa_i \psi_{i}} e^{-\omega_{i} \psi_{i}^2 / 2} \propto e^{\kappa_i \varphi_{i}} e^{ - \eta_\epsilon \omega_{i} \varphi_{i}} e^{-\omega_{i} \varphi_{i}^2 / 2}
\end{array}$ with $\varphi_i = \mathbf{m}_{i}^{\top} \boldsymbol{\eta}$. The conditional distributions, after applying the Esscher transform to both the PG and Gaussian variables, are given as
\begin{equation} \label{mcmc0}
\begin{array}{rl}
\left(\omega_i \mid-\right) &\sim \mathrm{PG}\left(y_i+\epsilon, \psi_i \right) \\
( \nu_i \mid - ) &\sim \mathrm{U} \{0, D_t \mathbf{z}^{\top}_{\alpha}(t_i) \mathbf{u}_{\alpha} \} \\
( \boldsymbol{\eta} \mid - ) &\sim \mathrm{TN} \left[ \left(\mathbf{M}^{\top} \boldsymbol{\Omega} \mathbf{M}\right)^{-1} \mathbf{M}^{\top} ( \boldsymbol{\kappa} - \eta_\epsilon \boldsymbol{\omega} ), \left(\mathbf{M}^{\top} \boldsymbol{\Omega} \mathbf{M}\right)^{-1}, \mathcal{C}_\nu \bigcap \mathcal{C}_0 \right]
\end{array}
\end{equation}
where the truncated normal is constrained to set: $\mathcal{C}_\nu \bigcap \mathcal{C}_0$ and $
\mathcal{C}_\nu = \bigcap_{\{i:y_i=1\}}^N \left\{ \boldsymbol{\eta} :  \nu_i \leq D_t \mathbf{z}^{\top}_{\alpha}(t_i) \mathbf{u}_{\alpha} \right\}$. The truncated normal density is proportional to the Gaussian kernel multiplied by an indicator for set constraints: 
$\boldsymbol{\eta} \sim \mathrm{TN}(\boldsymbol{\mu}, \boldsymbol{\Sigma}, \mathcal{C} )$,
$$
\begin{array}{c}
\pi ( \boldsymbol{\eta} \mid \boldsymbol{\mu}, \boldsymbol{\Sigma}, \mathcal{C} )
=
\frac{\exp \left\{-\frac{1}{2}( \boldsymbol{\eta}-\boldsymbol{\mu})^{\top} \boldsymbol{\Sigma}^{-1}( \boldsymbol{\eta}-\boldsymbol{\mu})\right\}}
{c_1 ( \boldsymbol{\mu}, \boldsymbol{\Sigma}, \mathcal{C} )} 
\mathbb{I}( \boldsymbol{\eta} \in \mathcal{C} ) 
\propto 
{\exp \left\{-\frac{1}{2}( \boldsymbol{\eta}-\boldsymbol{\mu})^{\top} \boldsymbol{\Sigma}^{-1}( \boldsymbol{\eta}-\boldsymbol{\mu})\right\}} 
\mathbb{I}( \boldsymbol{\eta} \in \mathcal{C} ) 
\end{array}
$$
where $c_1 ( \boldsymbol{\mu}, \boldsymbol{\Sigma}, \mathcal{C} ) = \oint_{ \boldsymbol{\eta} \in \mathcal{C} } \exp \left\{-\frac{1}{2}( \boldsymbol{\eta}-\boldsymbol{\mu})^{\top} \boldsymbol{\Sigma}^{-1}( \boldsymbol{\eta}-\boldsymbol{\mu})\right\} d \boldsymbol{\eta}$. Here $\boldsymbol{\Omega}=\operatorname{diag}\left(\omega_1, \cdots, \omega_N\right)$, $\omega_i$ make up the vector $\boldsymbol{\omega}$ and $\kappa_i = (y_i-\epsilon)/2$ and make up the vector $\boldsymbol{\kappa}$. Note that convex set intersection preserves convexity (\citealt{boyd2004convex} Chapter 2), allowing for the exploitation of convex set constrained Gaussian sampling \citep{maatouk2016new}. At this point, we have shown how to incorporate hazard rate constraints into the general Gibbs sampler. The priors for $\boldsymbol{\eta}$ are flat and we outline hierarchical Gaussian structures later in the article.

\subsection{Sufficient statistics dimension reduction for efficient computation}

A naive slice sampler enforces a prohibitive number of $\sum_i y_i$ inequality constraints in $\mathcal{C}_\nu$ during Gaussian sampling. The slice sampler for nonparametric partition estimators \citep{nieto2002markov,kaplan1958nonparametric,ibrahim2001bayesian} presents a unique opportunity to collapse the linear operator of Markov transitions based on sufficient statistics. By using Riemann sum integration to construct the baseline log cumulative hazard $\mathbf{z}^{\top}_{\alpha}(t_i) \mathbf{u}_{\alpha}$, we the simplify $\mathcal{C}_\nu$ and auxiliary transitions $( \nu_i \mid - ) \sim \mathrm{U} \{0, D_t \mathbf{z}^{\top}_{\alpha}(t_i) \mathbf{u}_{\alpha} \}$. The new hazard rate is now constructed with non-overlapping histogram partitions, using delta functions $D_t z_{\alpha,j} (t_i) =  \delta_j (t_i) = \mathbb{I}( s_{j-1} \leq t_i < s_j)$ as its additive bases with $D_t \mathbf{z}^{\top}_{\alpha}(t_i) \mathbf{u}_{\alpha} = \sum_{j=1}^J D_t z_{\alpha,j} (t_i) u_{\alpha,j}$. A example of the basis functions $D_t \mathbf{z}^{\top}_{\alpha}(t_i)$ used to construct the histogram in Riemann sum integration for the \texttt{lung} dataset from \texttt{R} \texttt{survival}  is presented in Figure~\ref{fig:2}a \citep{loprinzi1994prospective,therneau2023package}. We can write matrix $D_t \mathbf{Z}_\alpha \in \mathbb{R}^{N \times J}$ with rows $D_t \mathbf{z}_{\alpha}^{\top}(t_i)$. Now the matrix becomes $D_t \mathbf{Z}_\alpha \in \{ 0, 1 \}^{N \times J}$ with induced norm $\|D_t \mathbf{Z}_\alpha\|_{\infty} = 1$. Each row only has at most a single element that is 1, meaning each uncensored subject activates at most a single delta function. The uniform auxiliary space becomes the last order statistics of uniform random variables or beta random variables. An example of inequalities $\nu_i \leq D_t \mathbf{z}^{\top}_{\alpha}(t_i) \mathbf{u}_{\alpha}$ are given as
$$
\begin{array}{rlllllll}
\vdots &\leq \vdots & \vdots & \vdots & \vdots & \vdots & \vdots & \vdots\\
\nu_3 &\leq 0 &+ 0 &+ u_{\alpha,3} &+ 0 &+ 0 &+ 0 &+ 0\\
\nu_4 &\leq 0 &+ 0 &+ u_{\alpha,3} &+ 0 &+ 0 &+ 0 &+ 0\\
\nu_5 &\leq 0 &+ 0 &+ 0 &+ u_{\alpha,4} &+ 0 &+ 0 &+ 0\\
\vdots &\leq \vdots & \vdots & \vdots & \vdots & \vdots & \vdots & \vdots 
\end{array}
$$
and we see that for the set $\{i: y_i=1\}$ we have inequalities for each coefficient $u_{\alpha,j}$ multiplied together as indicators, 
$
\begin{array}{c}
\prod_{ \{ i: y_i=1 \} } \mathbb{I} \left\{ \delta_j (t_i) \nu_i \leq {u}_{\alpha,j} \right\}
\end{array}
$
and our set $\mathcal{C}_\nu$ consist of\\
$
\begin{array}{c}
\max \left\{ \delta_j (t_i) \nu_i :  y_i=1 \right\} \leq {u}_{\alpha,j}
\end{array}.
$
We denote $\nu_{i,j} = \delta_j (t_i) \nu_i$ and we only need to sample from the last order statistic 
$
\begin{array}{c}
v_j = \max \{ \nu_{i,j} : y_i=1, \delta_j(t_i)=1 \}.
\end{array}
$
Each sample of the uniform variable becomes $\nu_i \mid \{ \delta_j(t_i) = 1 \} \sim \mathrm{U} (0, u_{\alpha,j})$. Let the number of uncensored events in partition $j$ be $n_{\alpha,j} = \sum_{ i=1 }^N y_i \delta_j(t_i)$ and $V_j = \nu_{(n_{\alpha,j})}$ denote the last order statistic. We only need to use the last order statistic, the sufficient statistics of $\mathrm{U} (0, u_{\alpha,j})$ to construct $\mathcal{C}_\nu$ with reduced dimensions. The conditional cumulative distributions and densities for the last order statistics are
$$
\begin{array}{c}
\mathbb{P} \left( V_j \leq v_j \mid u_{\alpha,j}, n_{\alpha,j} \right) = \mathbb{I}(v_j \leq u_{\alpha,j}) \left( \frac{v_j}{u_{\alpha,j}} \right)^{n_{\alpha,j}},
\pi \left( v_j \mid u_{\alpha,j}, n_{\alpha,j} \right) = \mathbb{I}(v_j \leq u_{\alpha,j}) { n_{\alpha,j} v_j^{n_{\alpha,j}-1} }/{u_{\alpha,j }^{n_{\alpha,j}} }
\end{array}
$$
with $v_j \mid u_{\alpha,j}, n_{\alpha,j} \sim \mathrm{B} (u_{\alpha,j}, n_{\alpha,j}, 1)$ and $\mathcal{C}_v = \bigcap_{j=1}^J \left\{ \boldsymbol{\eta} :  v_j \leq {u}_{\alpha,j} \right\}$. This is a simple half space $\mathbf{v} \leq \mathbf{u}_\alpha$ and a important improvement over naive slice sampling strategies. Notably, we can rescale beta variables to sample from $\mathrm{B} (u_{\alpha,j}, n_{\alpha,j}, 1)$, where $(v_j/u_{\alpha,j}) \sim \mathrm{Beta}(n_{\alpha,j}, 1)$ can be verified using a Jacobian and has the property $0 \leq v_j/u_{\alpha,j} \leq 1$. The denominator terms $({u_{\alpha,j}})^{n_{\alpha,j}}$ cancels our new sufficiently reduced hazard term $\prod_{i=1}^N \left\{ D_t \mathbf{z}_{\alpha}^{\top} (t_i) \mathbf{u}_\alpha \right\}^{y_i} = \prod_{i=1}^N \prod_{j=1}^J  {u}_{\alpha,j}^{y_i \delta_j(t_i) }= \prod_{j=1}^J ({u_{\alpha,j}})^{n_{\alpha,j}}$. The new NB likelihood with beta kernels is given as $$
\begin{array}{c}
L_{\text{NB}}(\boldsymbol{\eta} \mid \mathbf{y}, \mathbf{M} ) \propto \left[ \prod_{i=1}^N \prod_{j=1}^J  {u}_{\alpha,j}^{y_i \delta_j(t_i) } \right]
\mathbb{I}( \boldsymbol{\eta} \in \mathcal{C}_0 ) \left[ \prod_{i=1}^N (p_i)^{y_i } (1 - p_i)^{\epsilon } \right]
\end{array}.
$$
The conditional distributions replacing equations \eqref{mcmc0} are now given as
\begin{equation} \label{mcmc1}
\begin{array}{rl}
\left(\omega_i \mid-\right) &\sim \mathrm{PG}\left(y_i+\epsilon, \psi_i \right) \\
\left( v_j \mid-\right) &\sim \mathrm{B} (u_{\alpha,j}, n_{\alpha,j}, 1) \\
( \boldsymbol{\eta} \mid - ) &\sim \mathrm{TN} \left\{ \left(\mathbf{M}^{\top} \boldsymbol{\Omega} \mathbf{M}\right)^{-1} \mathbf{M}^{\top} ( \boldsymbol{\kappa} - \eta_\epsilon \boldsymbol{\omega} ), \left(\mathbf{M}^{\top} \boldsymbol{\Omega} \mathbf{M}\right)^{-1}, \mathbf{v} \leq \mathbf{u}_\alpha \right\}.
\end{array}
\end{equation}
Note that $\{\mathbf{v} \leq \mathbf{u}_\alpha\} = \mathcal{C}_v \ne \mathcal{C}_v \bigcap \mathcal{C}_0$ if only censored events are found in a given partition. But we recommend to using a parametrization where $\{\mathbf{v} \leq \mathbf{u}_\alpha\} = \mathcal{C}_v \bigcap \mathcal{C}_0$ by ensuring each partition has at least one uncensored event. Under this construction, the spline $z_{\alpha,j}(t)$ is a monotonic step function with a 45 degree slope in the partition $s_{j-1} \leq t < s_j$ and is visualized in Figure \ref{fig:2}b. The slice constraints are now based on number of coefficients, instead of number of subjects. Many methods exist that can efficiently sample a Gaussian distribution constrained to $\{\mathbf{v} \leq \mathbf{u}_\alpha\}$, constraints reduced from number of subject $N$ to number of coefficients in $\mathbf{u}_\alpha$ \citep{valeriano2023moments}. Note that $\boldsymbol{\Omega}$ is a diagonal matrix and matrices $ \mathbf{M}^{\top} \boldsymbol{\Omega} \mathbf{M} $ and $ \mathbf{M}^{\top} ( \boldsymbol{\kappa} - \eta_\epsilon \boldsymbol{\omega} ) $ can be computed in a parallel manner without loading all of the data in memory \citep{polson2013bayesian}. In addition, we can further simplify sampling of $\boldsymbol{\eta} = \left[ \mathbf{u}^\top_{\alpha}, \boldsymbol{\beta}^\top \right]^\top$ through conditional Gaussian distributions,
$$
\begin{array}{c}
\underbrace{ \left( \mathbf{u}_\alpha \mid - \right) \sim \operatorname{TN} \left( \boldsymbol{\mu}_{u \mid \beta}, \Sigma_{u \mid \beta}, \textbf{v} \leq \textbf{u}_\alpha \right) }_{\text{baseline log cumulative hazard local slopes}}, \quad
\underbrace{ \left( \boldsymbol{\beta} \mid - \right)  \sim \operatorname{N}(\boldsymbol{\mu}_{\beta \mid u}, \Sigma_{\beta \mid u} ) }_{\text{PH coefficients \& intercept}}.
\end{array}
$$
A multilevel Gaussian model can be specified for $\boldsymbol{\beta}$ while a flat prior can be maintained for $\mathbf{u}_\alpha$. The interpretation of spline coefficients $\mathbf{u}_\alpha$ are positive  slopes for local linear interpolation of the baseline log cumulative hazard function to ensure a final monotonic function.

As we increase number of partitions $J$, we achieve a more accurate estimate at the expense of computation. This improves on standard nonparametric estimators \citep{kaplan1958nonparametric,nelson1972theory,aalen1978nonparametric,nieto2002markov} by using a continuous monotonic local linear regression to approximate the log cumulative hazard, rather than an empirical jump process. These splines are centered at their midpoint to reduce attenuation due to priors that can be incorporated (Figure \ref{fig:2}b). The intercept term $\alpha_0$ allows location shifts and is unconstrained under Cox regression. The truncated normal Gibbs step is an iterative sampling of non-negative local slopes. Under Bayesian inference, we condition the slope of each partition on the remaining variables. This is a variant of sufficient dimension reduction techniques \citep{li2018sufficient,sun2019counting} and these local slopes are an analog of the martingale increments \citep{tsiatis2006semiparametric} used to construct the partial likelihood under stochastic calculus \citep{kaplan1958nonparametric,cox1972regression,cox1975partial}.

\subsection{Case weights under sufficient statistics}
By reducing computation to beta kernels, case weights $w_i$ can easily be incorporated into the Gibbs sampler. The NB likelihood becomes 
$$
\begin{array}{c}
L_{\text{NB}}(\boldsymbol{\eta} \mid \mathbf{w}, \mathbf{y}, \mathbf{M} ) \propto \left[ \prod_{i=1}^N \prod_{j=1}^J  {u}_{\alpha,j}^{y_i \delta_j(t_i) w_i} \right]
\mathbb{I}( \boldsymbol{\eta} \in \mathcal{C}_0 ) \left[ \prod_{i=1}^N (p_i)^{y_i w_i} (1 - p_i)^{\epsilon w_i} \right]
\end{array}
$$ and beta distribution parameter is modified to be $n^\star_{\alpha,j} = \sum_{ i=1 }^N y_i \delta_j(t_i) w_i$. The updated Gibbs sampler is 
\begin{equation}\label{mcmc1.1}
\begin{array}{rl}
\left(\omega_i \mid-\right) &\sim \mathrm{PG}\left\{ (y_i+\epsilon) w_i, \psi_i \right\} \\
\left( v_j \mid-\right) &\sim \mathrm{B} (u_{\alpha,j}, n^\star_{\alpha,j}, 1) \\
( \boldsymbol{\eta} \mid - ) &\sim \mathrm{TN} \left\{ \left(\mathbf{M}^{\top} \boldsymbol{\Omega} \mathbf{M}\right)^{-1} \mathbf{M}^{\top} ( \boldsymbol{\kappa}^\star - \eta_\epsilon \boldsymbol{\omega} ), \left(\mathbf{M}^{\top} \boldsymbol{\Omega} \mathbf{M}\right)^{-1}, \mathbf{v} \leq \mathbf{u}_\alpha \right\}
\end{array}
\end{equation}
with vector $\boldsymbol{\kappa}^\star$ consisting of values $(y_i-\epsilon)w_i/2$. This allows the use of Cox-PG in weighted analysis settings such as power priors \citep{ibrahim2000power,ibrahim2015power} and inverse probability weighting \citep{shu2021variance}.

\subsection{General hierarchical Cox model: the Cox-PG algorithm}

Our previous Gibbs algorithm can sample from the Cox model, while mixed models for clustered data are straightforward extensions. We allow for additional heterogeneity through Gaussian random effects also known as log-normal frailties. We can also incorporate prior information on the baseline hazard and coefficients by specifying the appropriate Gaussian priors and formulating a general Gibbs sampler. We update the notation for $\boldsymbol{\eta}$ to include random effects and specify truncated Gaussian prior $\boldsymbol{\eta} \sim \mathrm{TN}( \mathbf{b}, \mathbf{A}(\tau_{\mathcal{B}})^{-1}, \mathcal{C}_0 )$ for coefficients given by the form $\mathbf{m}_{i}^{\top} \boldsymbol{\eta} = \mathbf{z}^{\top}_{\alpha}(t_i) \mathbf{u}_{\alpha} + \mathbf{x}^\top_i \boldsymbol{\beta} + \mathbf{z}^\top_{i,\mathcal{B}} \mathbf{u}_\mathcal{B}$. Let $\mathbf{A}( \tau_\mathcal{B} )^{-1}$ be the covariance matrix. We order $\mathbf{M}=[ \mathbf{Z}_\alpha, \mathbf{X}, \mathbf{Z}_\mathcal{B} ] \in \mathbb{R}^{N \times (J+P+M)}$, $\boldsymbol{\eta}=[ \mathbf{u}^{\top}_\alpha, \boldsymbol{\beta}^{\top}, \mathbf{u}^{\top}_\mathcal{B}]^{\top}$. Note that the intercept is combined into $\mathbf{X}$. Then $\mathbf{A}( \tau_{\mathcal{B}} ) = \boldsymbol{\Sigma}^{-1}_0 \oplus \tau_{\mathcal{B}} \mathbf{I}_{M} = \boldsymbol{\Sigma}^{-1}_\alpha \oplus \boldsymbol{\Sigma}^{-1}_x \oplus \tau_{\mathcal{B}} \mathbf{I}_{M}$ is the prior precision represented with direct sum operator $\oplus$ and $\boldsymbol{\Sigma}^{-1}_x$ is the precision for $\boldsymbol{\beta}$ and $\boldsymbol{\Sigma}^{-1}_\alpha$ is the precision of $\mathbf{u}_\alpha$ and $M$ is the number of random effect clusters for $\mathbf{Z}_\mathcal{B} \in \mathbb{R}^{N \times M} $. The prior mean is $\mathbf{b} = \left[ \boldsymbol{\mu}^\top_\alpha, \boldsymbol{\mu}^\top_x, \mathbf{0}_{M}^\top \right]^{\top}$ and $[ \mathbf{Z}_\alpha, \mathbf{X}] \in \mathbb{R}^{N \times (J+P)}$ are fixed effect covariates. Let $\mathcal{C}_0$ be a set of convex constraints for fixed effects $\{ \mathbf{u}_\alpha, \boldsymbol{\beta} \}$ only. The initial constraints are $\mathcal{C}_0 = \left\{ \boldsymbol{\eta} :  \mathbf{u}_{\alpha} > \mathbf{0} \right\}$ and we condition on $\epsilon$ through the offset term $\eta_{\epsilon}$ to model the baseline log cumulative hazard. We introduce the random effect structure by placing a gamma distribution on the precision: $\tau_\mathcal{B} \sim \mathrm{G} \left(a_0, b_0 \right)$, $\pi(\tau_\mathcal{B} \mid a_0, b_0) \propto \tau_\mathcal{B}^{a_{0}-1} \exp \left(-b_{0} \tau_\mathcal{B} \right) $. The Gibbs sampler for \eqref{mcmc0}, after applying sufficient dimension reduction and multiplying by the Gaussian and gamma priors, is given as 
\begin{equation} \label{mcmc2}
\begin{array}{rl}
\left( \omega_i \mid - \right) &\sim \mathrm{PG}\left(y_i+\epsilon, \psi_i \right) \\
\left( v_j \mid - \right) &\sim \mathrm{B} (u_{\alpha,j}, n_{\alpha,j}, 1) \\
\left( \tau_{\mathcal{B}} \mid - \right) &\sim \mathrm{G} \left( a_{0}+\frac{M}{2}, b_{0}+\frac{ \mathbf{u}^{\top}_\mathcal{B} \mathbf{u}_\mathcal{B}}{2} \right) \\
\left( \boldsymbol{\eta} \mid - \right) &\sim \mathrm{TN} \left( \mathbf{Q}^{-1} \boldsymbol{\mu}, \mathbf{Q}^{-1}, \mathbf{v} \leq \mathbf{u}_\alpha \right)
\end{array}
\end{equation}
with $\boldsymbol{\mu} = \mathbf{M}^{\top} ( \boldsymbol{\kappa} - \eta_\epsilon \boldsymbol{\omega} )  + \mathbf{A}(\tau_{\mathcal{B}}) \mathbf{b} = \mathbf{M}^{\top} ( \boldsymbol{\kappa} - \eta_\epsilon \boldsymbol{\omega} )  + [(\boldsymbol{\Sigma}^{-1}_0 \boldsymbol{\mu}_0)^\top, \mathbf{0}_M^\top]^\top$ and $\mathbf{Q} = \mathbf{M}^{\top} \boldsymbol{\Omega} \mathbf{M} + \mathbf{A}(\tau_{\mathcal{B}})$. We may also write $\boldsymbol{\mu} = \mathbf{M}^{\top} \boldsymbol{\kappa} - \mathbf{M}^{\top} \boldsymbol{\Omega} \mathbf{1} \eta_\epsilon  + [(\boldsymbol{\Sigma}^{-1}_0 \boldsymbol{\mu}_0)^\top, \mathbf{0}_M^\top]^\top = \widetilde{\boldsymbol{\mu}} - \mathbf{M}^{\top} \boldsymbol{\Omega} \mathbf{1} \eta_\epsilon $ where $\widetilde{\boldsymbol{\mu}} = \mathbf{M}^{\top} \boldsymbol{\kappa} + [(\boldsymbol{\Sigma}^{-1}_0 \boldsymbol{\mu}_0)^\top, \mathbf{0}_M^\top]^\top$. For brevity, we abbreviate this modeling approach as the Cox-PG algorithm. As noted in Remark 3.3 of \citealt{choi2013polya}, a proper posterior does not depend on a full rank design matrix. In addition, Bayesian inference using \eqref{mcmc2} is possible in settings with high dimensional covariates, $J+P>N$ \citep{polson2013bayesian,makalic2015simple}.

\section{Theoretical results: uniform ergodicity}\label{theory}
We establish uniform ergodicity for our Gibbs sampler and show that the moments of posterior samples for $\boldsymbol{\eta}$ exists which guarantees central limit theorem (CLT) results for posterior MCMC averages and consistent estimators of the associated asymptotic variance \citep{jones2004markov}. Our posterior inference of $\alpha(t)$ involves matrix multiplication of coefficients and basis functions, $\boldsymbol{\mathcal{Z}}_\alpha \mathbf{u}_\alpha$ or a decomposition of additive functions and have the same CLT properties. Here $\boldsymbol{\mathcal{Z}}_\alpha$ is the matrix representation of the monotonic spline bases over the range of the observed covariate data. Suppose $\boldsymbol{\eta}^{(n)}$ is a draw from the posterior distribution, then $\alpha^{(n)}(t) = \alpha_0^{(n)} + \boldsymbol{\mathcal{Z}}_\alpha \mathbf{u}^{(n)}_\alpha$ are posterior samples of nonparametric effects with MCMC averages of $\widehat{\alpha}(t) = \left( N_{\text{MC}} \right)^{-1} \sum^{N_{\text{MC}}}_{n=1} \alpha^{(n)}(t)$ with an example provided in Figure \ref{fig:2}c.

First, we show uniform ergodicity of $\boldsymbol{\eta}$ for our Gibbs sampler by taking advantage of the truncated gamma distribution \citep{choi2013polya,wang2018analysis} and slice sampler properties \citep{mira2002efficiency}. A key feature of this strategy is truncating the gamma distribution at a small $\tau_0$ results in useful inequalities related to ergodicity while allowing for a vague prior to be specified for $\tau_\mathcal{B}$. We denote the $\boldsymbol{\eta}$-marginal Markov chain as $\Psi \equiv \{ \boldsymbol{\eta}(n) \}^\infty_{n=0}$ and Markov transition density (Mtd) of $\Psi$ as 
$$
\begin{array}{c}
k\left( \boldsymbol{\eta} \mid \boldsymbol{\eta}^{\prime}\right)
=
\int_{\mathbb{R}_{+}^{J}} \int_{\mathbb{R}_{+}} \int_{\mathbb{R}_{+}^{N}} \pi( \boldsymbol{\eta} \mid \boldsymbol{\omega}, {\tau}_\mathcal{B}, \mathbf{v}, \mathbf{y}) \pi\left(\boldsymbol{\omega}, {\tau}_{\mathcal{B}}, \mathbf{v} \mid \boldsymbol{\eta}^{\prime}, \mathbf{y}\right) d \boldsymbol{\omega} d {\tau}_\mathcal{B} d \mathbf{v}
\end{array}
$$
where $\boldsymbol{\eta}^{\prime}$ is the current state, $\boldsymbol{\eta}$ is the next state. Space $\mathbb{R}_{+}^{N}$ is the PG support that contains $\boldsymbol{\omega}$, $\tau_{\mathcal{B}} \in \mathbb{R}_{+}$ and $\mathbf{v} \in \mathbb{R}_{+}^{J}$. In addition, placing upper bounds of the baseline hazard ensures finite integrals in the Markov transitions while accommodating vague priors. Under initial constraints $\mathcal{C}_0$, we see that $k \left( \boldsymbol{\eta} \mid \boldsymbol{\eta}^{\prime}\right)$ is strictly positive and Harris ergodic \citep{tierney1994markov,hobert2011data}. We show the Mtd of satisfies the following minorization condition: $k \left( \boldsymbol{\eta} \mid \boldsymbol{\eta}^{\prime}\right) \geq \delta h(\boldsymbol{\eta})$, where there exist a $\delta > 0$ and density function $h$, to prove uniform ergodicity (\citealt{roberts2004general}, \citealt{meyn2012markov} Chapter 16). Uniform ergodicity is defined as bounded and geometrically decreasing bounds for total variation distance to the stationary distribution in number of Markov transitions $n$, $\left\|K^{n}( \boldsymbol{\eta}, \cdot)-\Pi(\cdot \mid \mathbf{y})\right\|_{\text{TV}}:=\sup _{A \in \mathscr{B}}\left|K^{n}( \boldsymbol{\eta}, A)-\Pi(A \mid \mathbf{y})\right| \leq V r^{n}$ leading to CLT results (\citealt{van2000asymptotic} Chapter 2). Here $\mathscr{B}$ denotes the Borel $\sigma$-algebra of $\mathbb{R}^{P+M}$, $K(\cdot, \cdot)$ is the Markov transition function for the Mtd $k(\cdot, \cdot)$ 
$$
K\left( \boldsymbol{\eta}^{\prime}, A\right)=\mathbb{P}\left( \boldsymbol{\eta}^{(r+1)} \in A \mid \boldsymbol{\eta}^{(r)}=\boldsymbol{\eta}^{\prime}\right)=\int_{A} k\left( \boldsymbol{\eta} \mid \boldsymbol{\eta}^{\prime}\right) d \boldsymbol{\eta},
$$
and $K^{n}\left( \boldsymbol{\eta}^{\prime}, A\right)=\mathbb{P}\left( \boldsymbol{\eta}^{(n+r)} \in A \mid \boldsymbol{\eta}^{(r)}=\boldsymbol{\eta}^{\prime}\right)$.
We denote $\Pi(\cdot \mid \mathbf{y})$ as the probability measure with density $\pi(\boldsymbol{\eta} \mid \mathbf{y})$, $V$ is bounded above and $r\in(0,1)$.

\begin{enumerate}[label=(C\arabic*), ref=\arabic*]
\item \label{c1} The gamma prior parameters are constrained as $a_0 + M/2 \geq 1$, $b_0>0$ and gamma support is constrained to be $\tau_{\mathcal{B}} \geq \tau_0$.
\item \label{c2} The monotonic splines system are comprised of Lipschitz continuous basis functions with coefficients constrained to $\mathbf{u}_\alpha \in \left( \mathbf{0}, \mathbf{u}_\alpha^+ \right)$, such that $\mathcal{C}_0 \subseteq \{ \boldsymbol{\eta} : \mathbf{u}_\alpha \in \left( \mathbf{0}, \mathbf{u}_\alpha^+ \right) \}$ and $D_t \mathbf{z}_{\alpha}^{\top} (t_i) \mathbf{u}_\alpha < D_t \mathbf{z}_{\alpha}^{\top} (t_i) \mathbf{u}_\alpha^+<\infty$.
\end{enumerate}

\begin{theorem} \label{thm1}
Under conditions (C\ref{c1}) and (C\ref{c2}), the Markov chain \eqref{mcmc2} is uniformly ergodic.
\end{theorem}

\begin{theorem} \label{thm2}
Under conditions (C\ref{c1}) and (C\ref{c2}), for any fixed $\mathbf{t} \in \mathbb{R}^{P+M}$, $\int_{\mathbb{R}^{P+M}} e^{\boldsymbol{\eta}^{\top} \mathbf{t}} \pi(\boldsymbol{\eta} \mid \mathbf{y}) d \boldsymbol{\eta}<\infty$. Hence the moment generating function of the posterior associated with \eqref{mcmc2} exist.
\end{theorem}

We leave the proofs to the supplement. Condition (C\ref{c1}) can be disregarded when considering a Gibbs sampler without mixed effects; the Gibbs sampler is still uniformly ergodic under (C\ref{c2}). Condition (C\ref{c1}) bounds the second order variation of the random effects while allowing for vague priors through the truncated gamma support \citep{wang2018analysis}. Condition (C\ref{c2}) enforces distributional robustness \citep{blanchet2022confidence} on how much the baseline hazard can vary, with a broad class of vague and informative priors that can be accommodated. Specifically, it is a common condition that bounds the slopes of the log cumulative hazard function. In large sample theory, condition (C\ref{c2}) or bounding the baseline cumulative hazard is also used to facilitate the dominated convergence theorem and achieve strong consistency \citep{wang1985strong,mclain2013efficient}. This proof also guarantees uniform ergodicity for Poisson regression through NB approximations and PG samplers \citep{zhou2012lognormal,duan2018scaling}. The following results for coupling time $ \left\|K^{n}( \boldsymbol{\eta}, \cdot)-\Pi(\cdot \mid \mathbf{y})\right\|_{\text{TV}} \leq (1-\delta)^{n}$ is derived from the minorization condition \citep{jones2004markov}. Numerical integration can be used to calculate $\delta$ prior to MCMC sampling to evaluate convergence rates and determine number of MCMC samples $n$ \citep{jones2001honest,geweke1989bayesian}.
More importantly, these results establish conditions for convergence in total variation distance and posterior expectations for constrained nonparametric objects under Bayesian inference.

\section{Accelerated mixing and removing bias with calibration} \label{accel}
From Gibbs samplers, we have the useful property of reversible marginal transitions. This allows us to construct a simple Metropolis-Hasting correction to debias our MCMC estimate. The pure Poisson process version of this problem was studied in \cite{duan2018scaling} with slow mixing but accurate inference using $\epsilon=1000$. A large $\epsilon$ artificially induces an imbalanced logistic regression problem \citep{zens2023ultimate}. Data calibration can be used once a Gibbs sampler for approximate PH regression has been established \citep{duan2018scaling}. We can use Gibbs sampler \eqref{mcmc2} as a proposal distribution to sample from the target posterior constructed with the beta kernel Cox semiparametric regression posterior, $L_{\mathrm{PH}}(\boldsymbol{\eta} \mid \mathbf{y}, \mathbf{M} ) \pi_0(\boldsymbol{\eta})$ where $\pi_0(\boldsymbol{\eta})$ is the prior. We accept the new Gibbs draw from proposal distribution $q(\boldsymbol{\eta} \mid \boldsymbol{\eta}^\prime)$ using a MH step with probability 
\begin{equation} \label{acc1}
\begin{array}{c}
\min \left\{ 1, \frac{ L_{\mathrm{PH}}(\boldsymbol{\eta} \mid \mathbf{y}, \mathbf{M} ) \pi_0(\boldsymbol{\eta}) q\left( \boldsymbol{\eta}^{\prime} \mid \boldsymbol{\eta} \right)
}{ 
L_{\mathrm{PH}}(\boldsymbol{\eta}^\prime \mid \mathbf{y}, \mathbf{M} ) \pi_0(\boldsymbol{\eta}^\prime) q\left( \boldsymbol{\eta} \mid \boldsymbol{\eta}^{\prime}\right)} \right\}
=
\min \left\{1, \prod_{i=1}^N \frac{ \exp \left\{ \lambda_i^\prime \right\}}{ \exp \left\{ \lambda_i \right\}} \frac{\left\{1+\exp \left(\psi_i\right)\right\}^{y_i + \epsilon}}{\left\{1+\exp \left(\psi_i^\prime\right)\right\}^{y_i + \epsilon}}\right\}
\end{array}.
\end{equation}
We denote current state as $\psi_i^\prime = \log(\lambda_i^\prime) - \log(\epsilon) = \mathbf{m}_{i}^{\top} \boldsymbol{\eta}^\prime - \log(\epsilon)$ and next state as $\psi_i = \log(\lambda_i) - \log(\epsilon) = \mathbf{m}_{i}^{\top} \boldsymbol{\eta} - \log(\epsilon)$. The derivation of this probability can be done quickly by noting that Gibbs samplers are MH algorithms with acceptance probability one. Therefore under the Cox-PG Gibbs sampler for target posterior $L_{\mathrm{NB}}(\boldsymbol{\eta} \mid \mathbf{y}, \mathbf{M} ) \pi_0(\boldsymbol{\eta})$ we have
\begin{equation} \label{acc2}
\begin{array}{c}
\frac{ L_{\mathrm{NB}}(\boldsymbol{\eta} \mid \mathbf{y}, \mathbf{M} ) \pi_0(\boldsymbol{\eta}) q\left( \boldsymbol{\eta}^{\prime} \mid \boldsymbol{\eta} \right)
}{ 
L_{\mathrm{NB}}(\boldsymbol{\eta}^\prime \mid \mathbf{y}, \mathbf{M} ) \pi_0(\boldsymbol{\eta}^\prime) q\left( \boldsymbol{\eta} \mid \boldsymbol{\eta}^{\prime}\right)} = 1
\implies 
\frac{ 
q\left( \boldsymbol{\eta}^{\prime} \mid \boldsymbol{\eta} \right)
}{ 
q\left( \boldsymbol{\eta} \mid \boldsymbol{\eta}^{\prime}\right)
}
=
\frac{
L_{\mathrm{NB}}(\boldsymbol{\eta}^\prime \mid \mathbf{y}, \mathbf{M} ) \pi_0(\boldsymbol{\eta}^\prime)
}{
L_{\mathrm{NB}}(\boldsymbol{\eta} \mid \mathbf{y}, \mathbf{M} ) \pi_0(\boldsymbol{\eta})
}
\end{array}.
\end{equation}
Substituting \eqref{acc2} into \eqref{acc1}, we have a ratio of likelihoods which simplifies to approximately a ratio of survival functions or a difference of cumulants on the log scale. This nested detailed balance structure removes the bias of $\text{G}(\epsilon,\epsilon)$ convolution and allows moderate $\epsilon = 100$ to be used to accelerate mixing. The frailty model with $\epsilon = 100$ modifies the original Cox model to have more heterogeneity, making it an ideal proposal distribution because proposal distributions should have slightly more variation than their target distribution. Through our experiments, we found $\epsilon = 100$ and the MH step to work well with over 90\% acceptance rates in various settings. The bias due to frailty in Gibbs samplers \eqref{mcmc2} can be removed with MH acceptance probability defined in \eqref{acc1}. For example, draws of $\boldsymbol{\eta}$ are sampled sequentially from \eqref{mcmc2}, each new draw of $\boldsymbol{\eta}$ is accepted with probability \eqref{acc1}.  After removing the bias, we have draws from the PH likelihood based posterior and we can use the log--log link (i.e. $\exp[-\exp\{A(t_i)\}]$) to map to the survival probabilities.

For Cox-PG with weights \eqref{mcmc1.1}, the ratios \eqref{acc1} and \eqref{acc2} are calculated using weighted likelihoods $L_{\text{NB}}(\boldsymbol{\eta} \mid \mathbf{w}, \mathbf{y}, \mathbf{M} )$ and $L_{\text{PH}}(\boldsymbol{\eta} \mid \mathbf{w}, \mathbf{y}, \mathbf{M} )$,
$$
\begin{array}{c}
\min \left\{1, \prod_{i=1}^N \left[ \frac{ \exp \left\{ \lambda_i^\prime \right\}}{ \exp \left\{ \lambda_i \right\}} \frac{\left\{1+\exp \left(\psi_i\right)\right\}^{y_i + \epsilon}}{\left\{1+\exp \left(\psi_i^\prime\right)\right\}^{y_i + \epsilon}} \right]^{w_i} \right\}
\end{array}.
$$
Aside from mixed models, all hierarchical models that can be computed with Gibbs transitions simplifies to convenient acceptance probability  \eqref{acc1}. These results are due to the prior $\pi_0(\boldsymbol{\eta})$ canceling out in the ratios.

\section{Other extensions of the Cox-PG algorithm}
\label{misc}

The framework we have established has wide applicability to numerous other contexts and problems. Any Gaussian Gibbs sampler model on the PH coefficients can be incorporated as direct extensions of our Cox-PG algorithm, including variable selection \citep{tipping2001sparse,makalic2015simple}. We detail a few additional extensions here:

\textit{Stratified hazards.} Two sets of baseline hazard spline systems can be specified based with interaction variables. For example, male and female interactions are given as $\mathbf{m}_{i}^{\top} \boldsymbol{\eta} = \alpha_{\text{male}} + \mathbf{z}^{\top}_{\alpha}(t_i) \mathbf{u}_{\alpha,\text{male}} + \alpha_{\text{female}} + \mathbf{z}^{\top}_{\alpha}(t_i) \mathbf{u}_{\alpha,\text{female}} + \mathbf{x}^\top_i \boldsymbol{\beta}$.

\textit{Smoothing splines as random effects.} Smoothing splines can also be modeled with random effects summarized in \citealt{wand2008semiparametric,o1986statistical,lee2018bayesian}. Similar to generalized additive models (GAMs), a nonparametric smooth effect of a continuous covariate can be obtain through penalized regression coefficients. The O’Sullivan spline approach found in \cite{wand2008semiparametric} uses a linear fixed effect and L2 penalized coefficients on Demmler-Reinsch (DR) oscillation basis functions \citep{demmler1975oscillation}. This conveniently corresponds with the second order derivative smoothness penalty and can be fitted with random effects using \eqref{mcmc2}.

\textit{Power prior for incorporating historical clinical trial data.} The power prior of \cite{ibrahim2000power} represents $L(\boldsymbol{\eta} \mid \mathbf{y}^\dagger, \mathbf{M}^\dagger)$ as the historical data prior where $a \in [0,1]$ is a hyper-parameter for including historical data. Under power prior form, we evaluate $L(\boldsymbol{\eta} \mid \mathbf{y}, \mathbf{M} ) \left[ L(\boldsymbol{\eta} \mid \mathbf{y}^\dagger, \mathbf{M}^\dagger) \right]^a$ \citep{ibrahim2015power}. This results in $a$ being multiplied with $y_i^\dagger$ and $\epsilon$ resulting in Cox-PG with weights, \eqref{mcmc1.1}.

\section{Illustrative example: KM curve}
\label{examples}

\begin{figure}[H]
\centering
\includegraphics[width=5in]{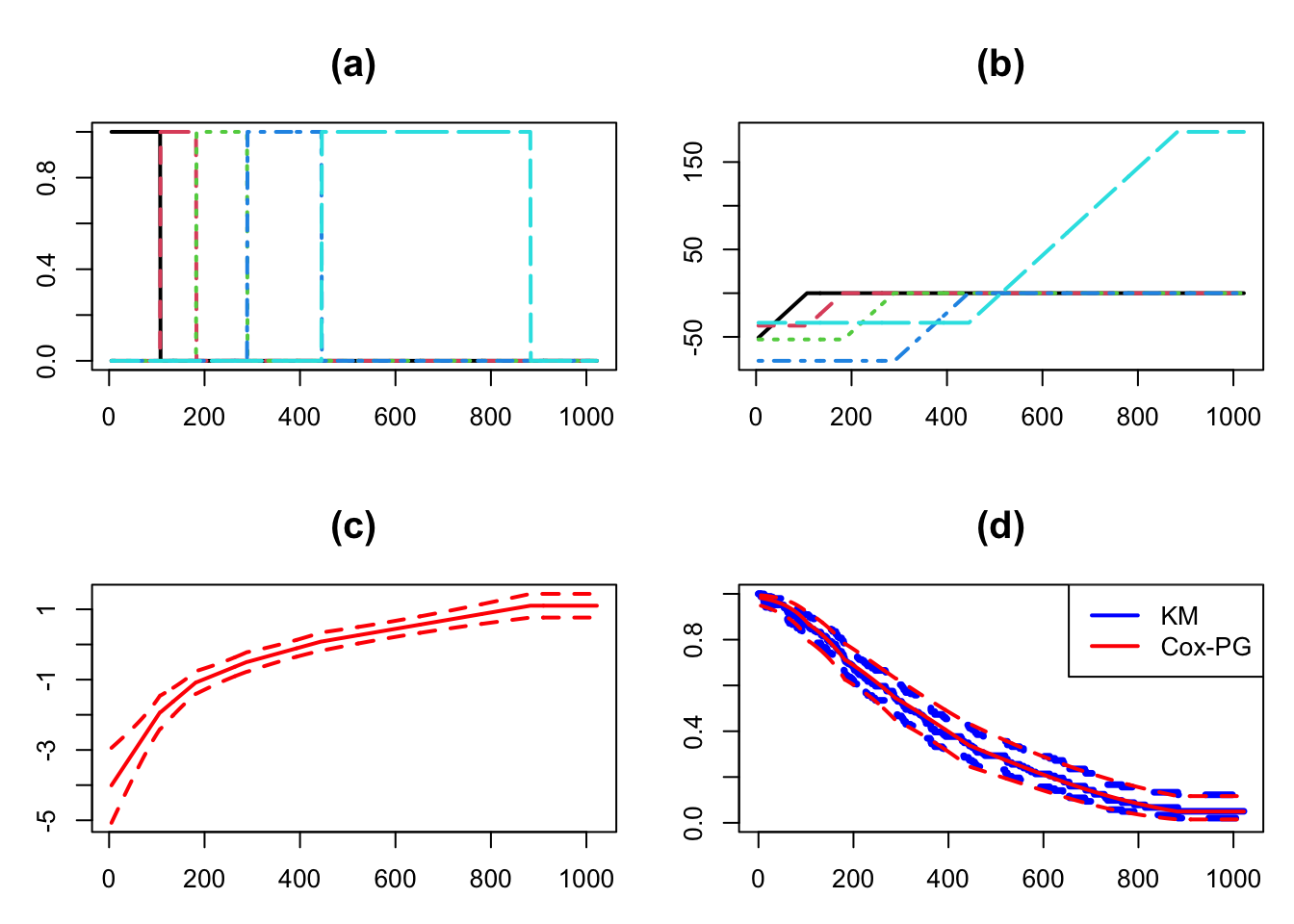}
{\linespread{1}\selectfont
\caption{Kaplan-Meier example with lung dataset. (a) Derivative of monotonic spline system $D_t \boldsymbol{\mathcal{Z}}_\alpha$ resulting in non-overlaping delta functions with likelihood contribution as beta kernels: $\prod_{i=1}^N \left\{ D_t \mathbf{z}_{\alpha}^{\top} (t_i) \mathbf{u}_\alpha \right\}^{y_i} = \prod_{i=1}^N \prod_{j=1}^J  {u}_{\alpha,j}^{y_i \delta_j(t_i) }$. (b) Monotonic spline system $\boldsymbol{\mathcal{Z}}_\alpha$ centered at midpoints. (c) Cox-PG posterior mean and 0.95 joint bands for baseline log cumulative hazard function. (d) Cox-PG and KM estimates for survival function.}\label{fig:2}}
\end{figure}

We apply Cox-PG to model survival curves using the \texttt{lung} dataset from the \texttt{survival} \texttt{R} package for illustrative purposes \citep{loprinzi1994prospective,therneau2023package}. Naturally, we do not need many partitions to fit a survival curve. Partitions boundaries $s_j$ should be place near the few changes in the second derivatives of the log cumulative hazard. We found that $J=5$ partitions works well with boundaries $s_j$ selected as equally spaced quantiles of uncensored times $\{t_i:y_i=1\}$ in the same manner as knot selection in functional data analysis \citep{ruppert2003semiparametric,ramsay2005fitting}. This also results in $n_{\alpha,1} \approx n_{\alpha,2}\approx \dots \approx n_{\alpha,J}$, equal numbers of uncensored events in each partition (Figure \ref{fig:2}a). This is an intuitive selection of partition boundaries as deaths drive the rate of survival processes. In addition, this spline-based approach enforces a horizontal line if the last set of study times $t_i$ are all censored events, similar to Kaplan-Meier (KM) estimates and is demonstrated in Figure \ref{fig:2}c \& d. This horizontal region coincides with the cessation of martingale calculus in KM and partial likelihoods (Figure \ref{fig:2}d). Using $\epsilon=100$, $J=5$ partitions and MH step \eqref{acc1} to remove bias, we achieve fast computation and accurate results. For the truncated Gaussian sampling, we use recently developed efficient Gibbs sampler package \texttt{relliptical} \citep{valeriano2023moments}. We use the \texttt{BayesLogit} package for PG sampling \citep{polson2013bayesian} and set $\boldsymbol{\Sigma}_0 = 10^6 \mathbf{I}$, $\mathbf{b} = \mathbf{0}$ as default parameters. In the back-end, study times are scaled and registered on compact interval $\left[ 0, 0.5 \right]$ in order to mitigate numerical underflow due to numerous inner product calculations. The code is provided in the supplement with convenient \texttt{R} function \texttt{posterior\_draw()} for posterior sampling. We drew $1000$ burn-in samples and then drew $100000$ posterior samples and thinned to retain $1000$ samples. This process was very fast and on average, took 2--3 minutes on an Apple M2 Pro Mac mini, 16GB. 

We map the Cox-PG posterior mean baseline log cumulative hazard $\widehat{\alpha}(t)+\widehat{\alpha}_0$ (Figure \ref{fig:2}c) to survival curves, using $\exp\{-\exp[\widehat{\alpha}(t)+\widehat{\alpha}_0]\}$ and compared it to Kaplan-Meier estimates (Figure \ref{fig:2}d). We plot our monotone basis functions $\boldsymbol{\mathcal{Z}}_\alpha$ for $J=5$ in Figure \ref{fig:2}b. We note that the posterior mean of the baseline log cumulative hazard is monotonic because it is a sum of monotonic functions and every posterior draw is a monotonic function. We use the approach of \citealt{meyer2015bayesian} to construct $0.95$ multiplicity adjusted joint bands for $\alpha(t)$. Our bounds are mapped to survival probabilities and are plotted in Figure \ref{fig:2}d. 

A novel and key advantage of our approach is that Cox-PG enforces monotonicity and continuity at every MCMC draw as local linear regression of the log cumulative hazards. This also induces a degree of smoothness in the survival curve, a useful safeguard against over fitting. In addition, fit can be improved with better partition selection and larger $\epsilon$. The KM example is a special case of Cox-PG with intercept $\alpha_0$ being the only unconstrained PH regression coefficient. We present PH regression models with different choices of $J$ and $\epsilon$ next.

\section{Simulation study: Weibull PH model}
\label{sim}
We used the following configurations of Cox-PG. \textit{Cox-PG1}: \eqref{mcmc2} without Metropolis-Hastings bias removal, $J=5$, $\epsilon=1000$. \textit{Cox-PG2}: \eqref{mcmc2} with Metropolis-Hastings bias removal, $J=5$, $\epsilon=100$. \textit{Cox-PG3}: \eqref{mcmc2} with Metropolis-Hastings bias removal, $J=10$, $\epsilon=100$. In addition, we compare our method with Bayesian integrated nested Laplace approximation (INLA) \citep{alvares2024bayesian,rustand2024fast,rue2009approximate}. As a gold standard method, we also fit the Weibull simulated data with a Bayesian Weibull PH model using a Hamiltonian Monte Carlo (HMC) algorithm built on \texttt{Stan} \citep{burkner2017brms}. Competing methods used default initialization conditions and parameters found in \texttt{R} package \texttt{INLAjoint} and \texttt{R} package \texttt{brms}. These methods are referred to as \textit{INLA} and \textit{HMC-Weibull} in the results.

As the basis of our simulation, we use the Weibull PH model $\alpha_1(t_i)+ x_{1i} \beta_1+ x_{2i} \beta_2$ where $\alpha_1(t_i) = \log(0.1)+2\log \left(t_i\right)$ is the baseline log cumulative hazard and $X_{1 i} \sim \mathrm{N}(0,1)$, $X_{2 i} \sim \operatorname{U}(-1,1)$ and $\{\beta_1=0.5,\beta_2=-0.5\}$. We simulate 4 cases, different configurations of the Weibull PH model. \textit{Frailty}: we simulate 25 balanced cluster random effects and add it to the Weibull process $\alpha_1(t_i)+ x_{1i} \beta_1+ x_{2i} \beta_2 +b_k$, with $b_k \sim \mathrm{N}(0,1)$. \textit{Weighting}: we first simulate the Weibull event times from $\alpha_1(t_i)+ x_{1i} \beta_1+ x_{2i} \beta_2$ and then contaminate 10\% of the data by replacing covariates with draws $X_{1 i} \sim \operatorname{U}(-10,10)$, $X_{2 i} \sim \operatorname{U}(-10,10)$; we assigned contaminated data a weight of 0.001 to mitigate contamination. \textit{GAM}: we add a nonparametric effect to the Weibull process $\alpha_1(t_i)+ x_{1i} \beta_1+ x_{2i} \beta_2 + \sin(x_{3i})$, with $x_{3i} \sim \mathrm{U}(0,2\pi)$. \textit{Stratified}: we sample 75\% of the data $\alpha_1(t_i)+ x_{1i} \beta_1+ x_{2i} \beta_2$ and 25\% of the data from $\alpha_2(t_i)+ x_{1i} \beta_1+ x_{2i} \beta_2$ with $\alpha_2(t_i)=\log(0.2)+\log \left(t_i\right)$; indicators for hazard groups are recorded. For each case, we sample $N=200$ observations and we drew independent censoring times from $\text{Exp}(0.1)$. For each case, we simulate 200 replicates and all were fitted with the 5 competing methods: Cox-PG1, Cox-PG2, Cox-PG3, INLA, and HMC-Weibull. HMC-Weibull can accommodate weights, frailties and GAMs but cannot fit stratified hazards or nonparametric baseline hazards. INLA can accommodate frailties, stratified hazards and nonparametric baseline hazards but cannot accommodate weights and GAMs. Cox-PG can accommodate the parameterization of all cases. All Cox-PG algorithms are initialized at the MLE and can be fitted with our \texttt{posterior\_draw()} function. All MCMC methods sampled 1000 burn-in and 10000 draws that were thinned to 1000 samples.

\begin{figure}[H]
\centering
\includegraphics[width=5in]{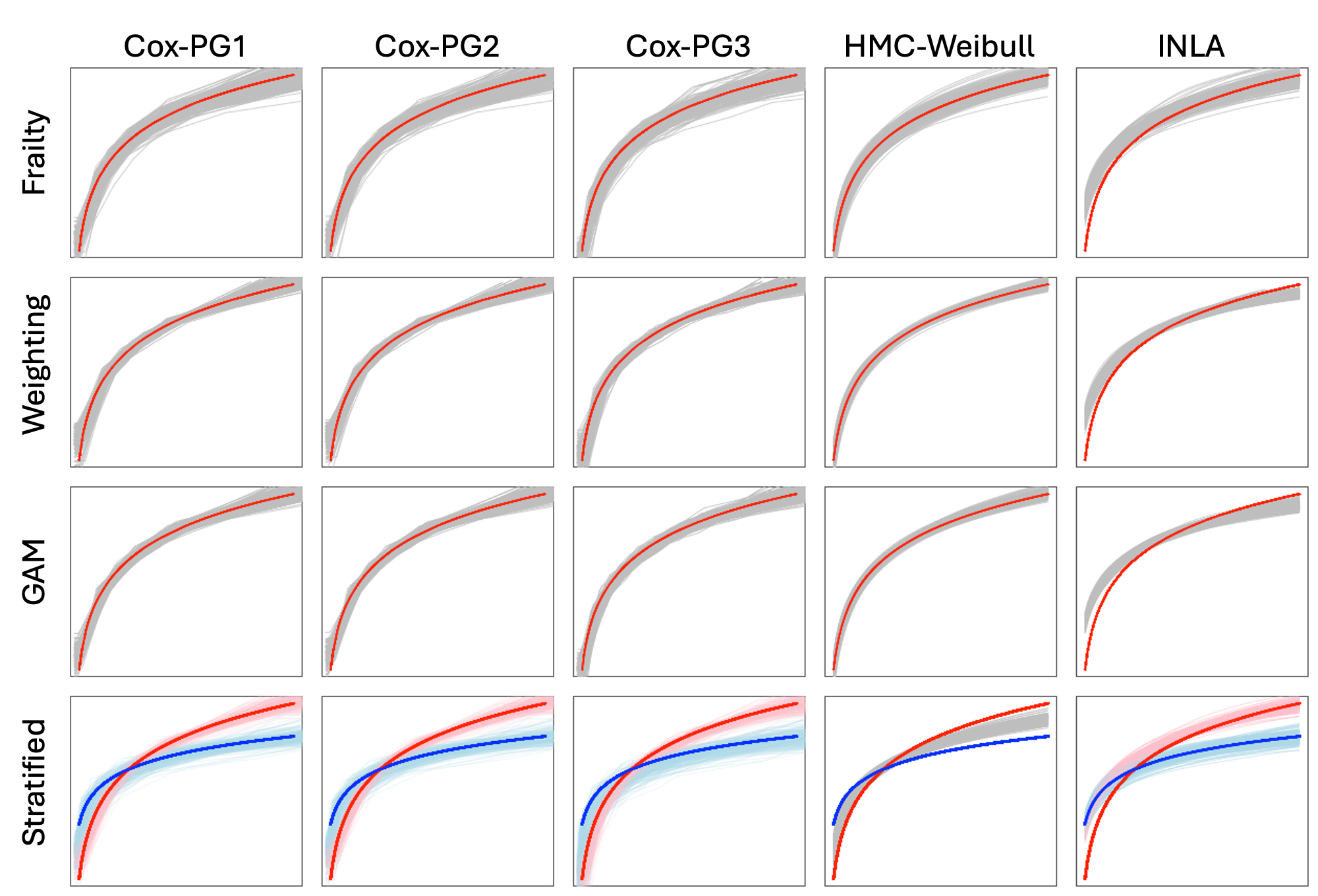}
{\linespread{1}\selectfont
\caption{Posterior estimates for Weibull log cumulative hazard $\alpha_1(t_i) = \log(0.1)+2\log \left(t_i\right)$. Estimates from $N=200$ replicates are plotted in gray. The true $\alpha_1(t_i)$ is plotted in solid red. For the \textit{Stratified} case $\alpha_2(t_i) = \log(0.2)+\log \left(t_i\right)$ is plotted in solid blue and estimates are plotted in transparent red and blue.}\label{fig:4}}
\end{figure}

For the baseline log cumulative hazard, we plot the integrated square error (ISE) $(U-L)^{-1}\int^U_L \left[ \alpha(t) - \widehat{\alpha}(t) \right]^2 dt $ and integrated coverage rate $(U-L)^{-1} \int^U_L \mathbb{I} \left[ \widehat{\alpha}^-(t) < \alpha(t) < \widehat{\alpha}^+(t) \right] dt$ in Figure \ref{fig:3}, where $\widehat{\alpha}^+(t)$ and $\widehat{\alpha}^-(t)$ are upper and lower bounds respectively. Integral domain $L$ and $U$ are set as the first and last death observed in the data. For HMC-Weibull and Cox-PG we were able to calculate the joint bands \citep{meyer2015bayesian} and we used the provided bounds from INLA. We plot estimates of $\widehat{\alpha}(t)$ based on posterior means in Figure \ref{fig:4}. Cox-PG has comparable coverage to HMC-Weibull which is specified under oracle knowledge of the true hazard family. The piecewise slope of Cox-PG estimates the shape of the baseline log cumulative hazard function well. INLA struggles using numerical integration to approximate the hazard function near the boundary $t \approx 0$ (Figure \ref{fig:4}). HMC-Weibull with knowledge of the true baseline hazard family outperforms Cox-PG in terms of  ISE. However, knowledge of the true hazard family is not possible in real data analysis. For the \textit{Stratified} case, INLA and Cox-PG outperforms HMC-Weibull by specifying two separate baseline hazards and the HMC-Weibull estimate bisects the two true hazard functions. Considering INLA uses $15$ partitions by default, Cox-PG outperforms INLA with fewer parameters in estimating $\alpha(t)$. For the \textit{GAM} case, results for estimating $\sin(x_3)$ can be found in the supplement.

\begin{figure}[H]
\centering
\includegraphics[width=5in]{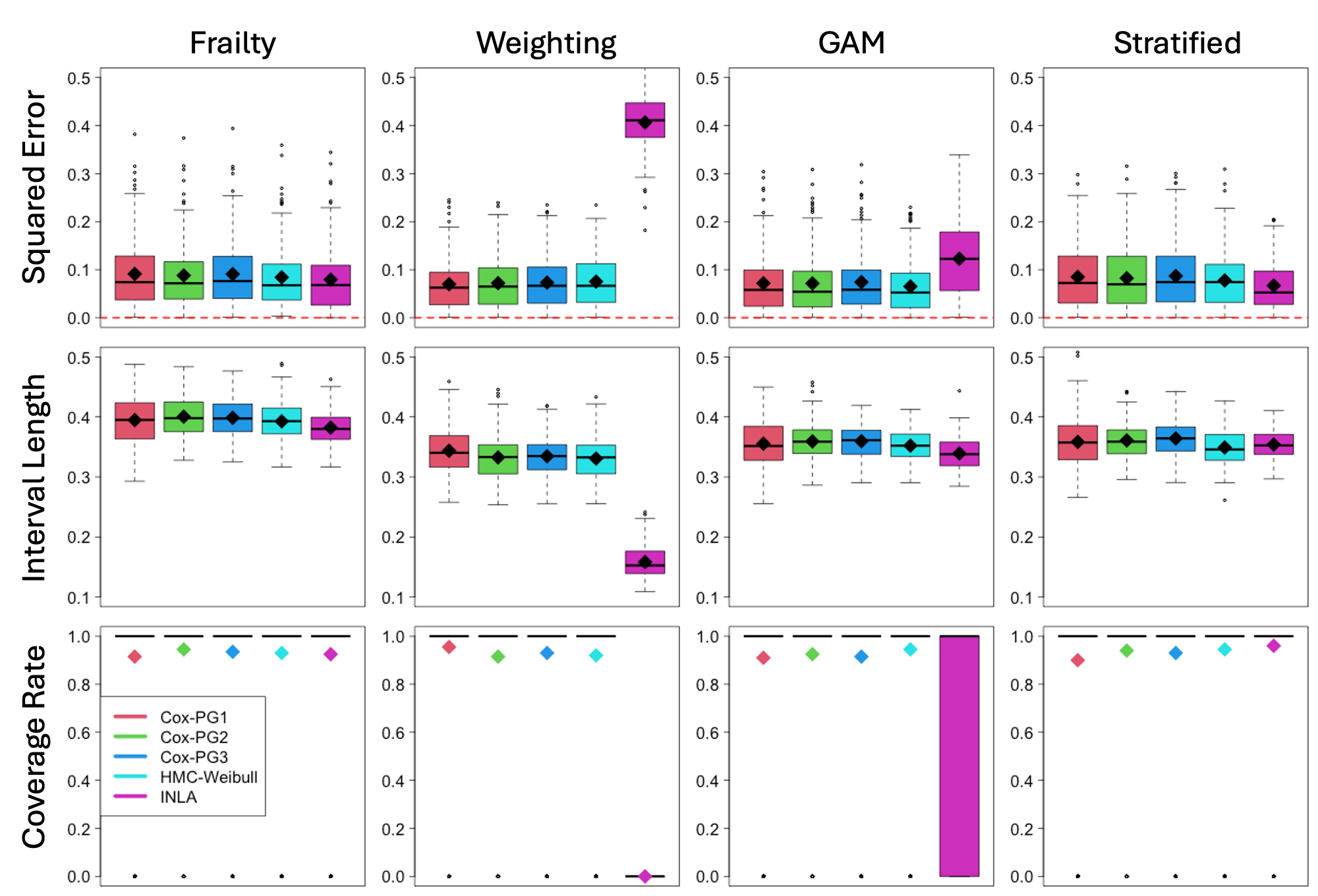}
{\linespread{1}\selectfont
\caption{Boxplot of squared errors, interval lengths and coverage rates for $\beta_1=0.5$. Means are denoted with diamond.}\label{fig:5}}
\end{figure}

\begin{figure}[H]
\centering
\includegraphics[width=5in]{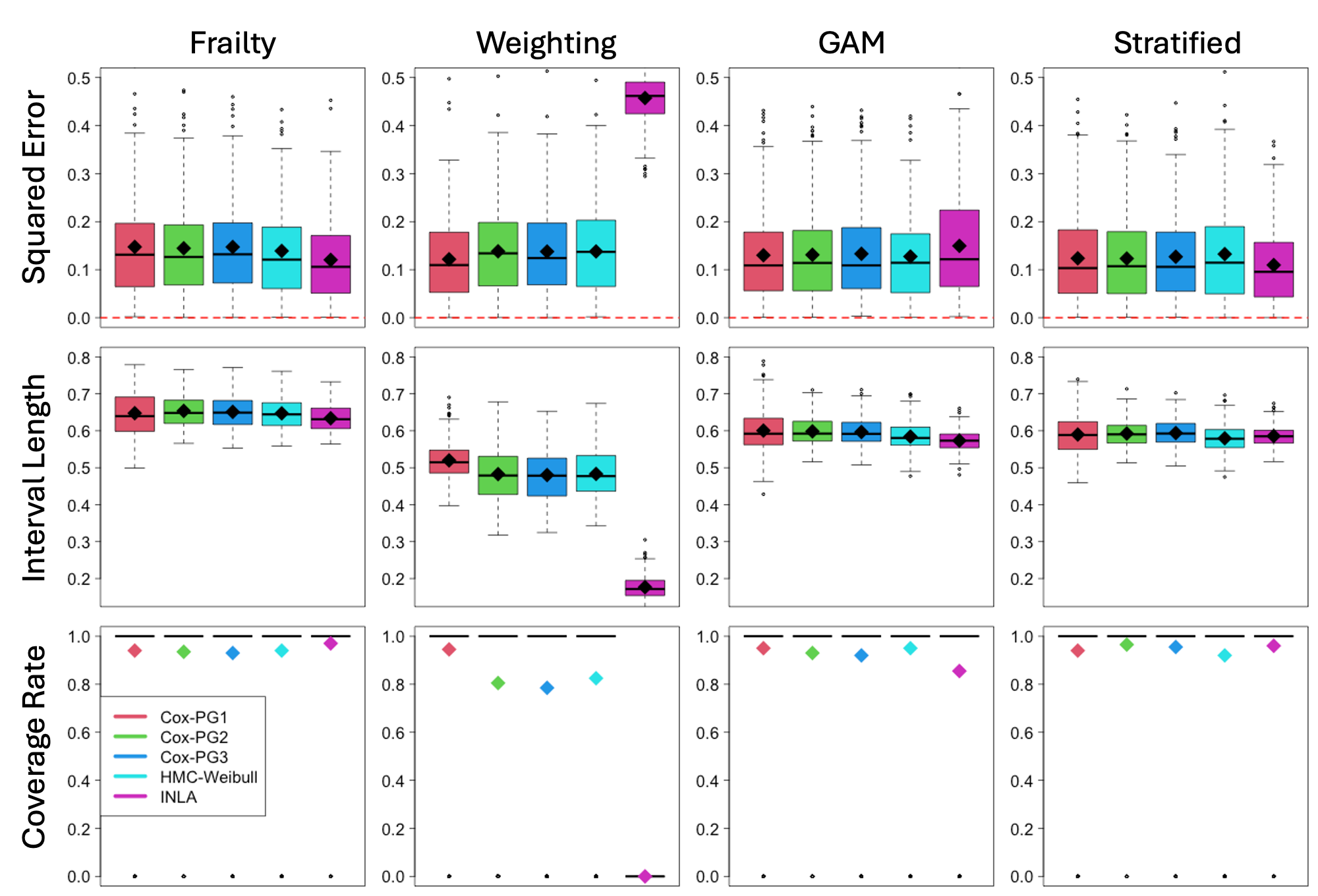}
{\linespread{1}\selectfont
\caption{Boxplot of squared errors, interval lengths and coverage rates for $\beta_2=-0.5$. Means are denoted with diamond.}\label{fig:6}}
\end{figure}

Squared errors, 0.95 credible interval length and 0.95 credible interval coverage rate for coefficient estimate of $\{\beta_1=0.5,\beta_2=-0.5\}$ are presented in Figure \ref{fig:5} \& \ref{fig:6}. All competing methods perform comparably in the \textit{Frailty} case, because all methods can fit mixed models. For analysis of both $\alpha(t)$ and $\boldsymbol{\beta}$, INLA performance suffers in cases where the weights and GAMs are not used in estimation. However, in cases of \textit{Frailty} and \textit{Stratified}, estimation of $\boldsymbol{\beta}$ remains accurate when there is a reasonable approximation of $\alpha(t)$; this robustness property is well documented in the literature \citep{tsiatis2006semiparametric,ibrahim2001bayesian}. Cox-PG1, without bias removal, slightly suffers in performance due to its additional heterogeneity and bias. As sample size increases, $n_{\alpha,j}$ increases leading to $v_j/u_{\alpha,j} \rightarrow 1$ from the left. This leads to slow mixing of the MCMC by causing $u_{\alpha,j}^{(n-1)} \approx u_{\alpha,j}^{(n)}$. We can decrease $n_{\alpha,j}$ by increasing the number of partitions $J$ at the cost of computation. However, Cox-PG3 can be unstable when there is not enough uncensored data to fit $J=10$ partitions, an identifiability problem, such as the \textit{Stratified} case with only 50 subjects in the $\alpha_2(t)$ group. Due to its flexibility, fewer monotonic spline parameters, bias removal and its comparable performance to gold standard methods, we recommend Cox-PG2 as the default.

\section{Leukemia example: stratified hazards, GAM, and frailties} \label{leuk}
We used the leukemia death/right-censored dataset from \cite{henderson2002modeling} with $N=1043$ and baseline covariates, found in \texttt{R} package \texttt{INLA} \citep{rue2017bayesian}. We include age and sex as linear PH effects. There are 24 district random effects denoting different sites. In addition, we account for nonlinear effect of the Townsend score: a quantitative measure with a range of $[-7,10]$ where higher values indicate less affluent areas. We believe that low scores $\approx -7$ have a greater association with increased survival that cannot be explained using a linear effect. The nonlinear GAM component induces an additional 7 random effects to model. We stratified the data into two baseline hazard groups based on white blood cell count to study the relationship between low count (leukopenia) and survival rates. We divided the data into two groups based on white blood cell count cutoff $<3 \times 10^9 / L$ for leukopenia and considered it normal otherwise. Cox-PG enables us to study the nonlinear effect of Townsend score and stratified baseline hazards, two violations of the standard proportional hazard assumption. For comparison, we fit Cox-PG2 and HMC-Weibull, using 1000 burn-in and 200000 draws that were thinned to retain 1000 samples. Our model has $M=31$ random effects, $P=5$ fixed effects/intercepts and $2J=10$ monotonic splines. We use our \texttt{posterior\_draw()} function for Bayesian computation.

\begin{figure}[H]
\centering
\includegraphics[width=5in]{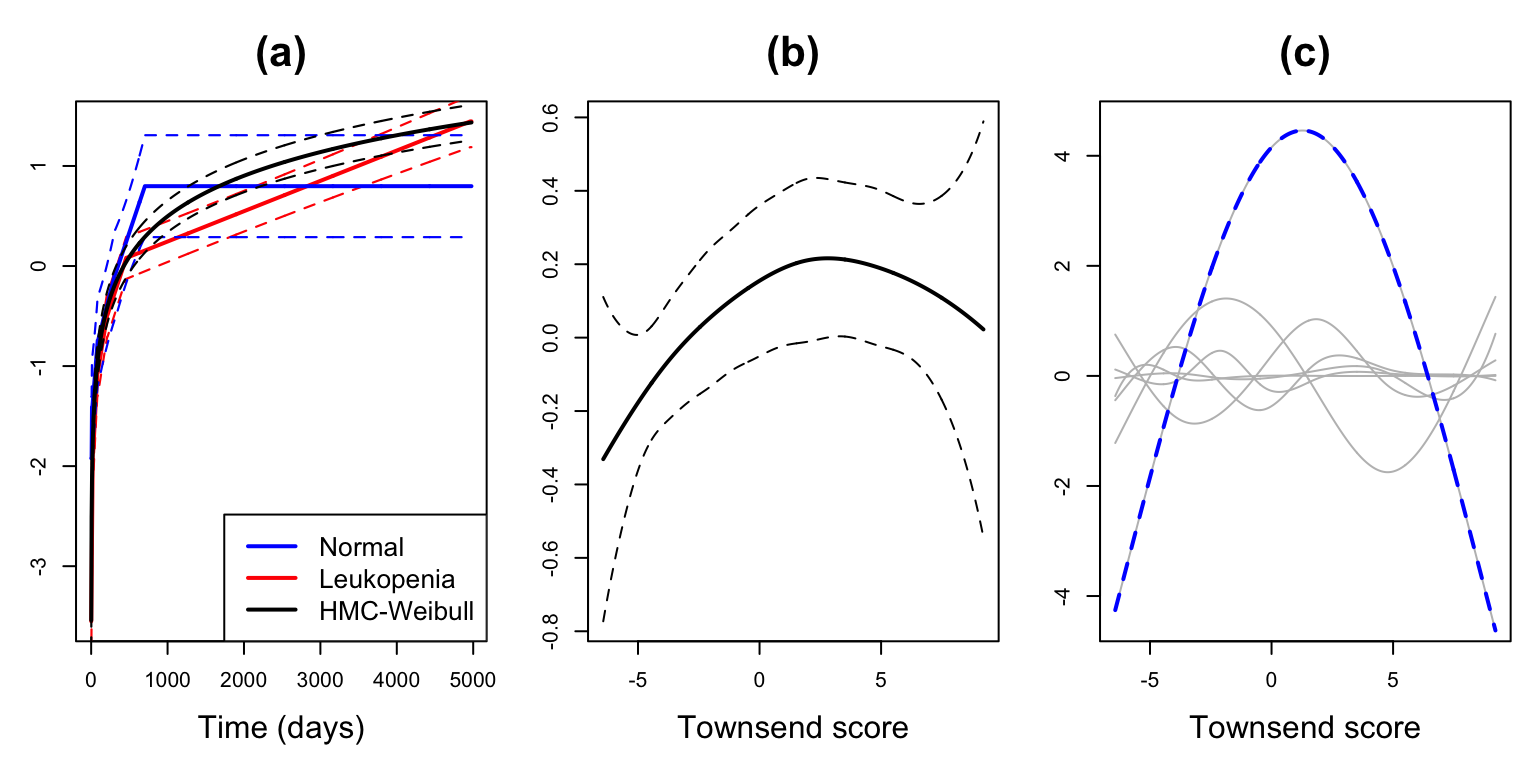}
{\linespread{1}\selectfont
\caption{Cox-PG estimates of baseline log cumulative hazard and Townsend effect from the leukemia dataset. (a) Estimates and joint bands for the stratified baseline log cumulative hazard along with the combined hazard from HMC-Weibull fit. (b) Smoothed nonparametric effect of Townsend score and 0.95 joint band. (c) Demmler-Reinsch (DR) basis system used to fit GAMs are plotted in gray; DR basis function corresponding to the coefficient with significant Bayesian p-value ($0.014$) is highlighted with a dashed blue line.}\label{fig:7}}
\end{figure}

HMC-Weibull can parameterize the site random effect and GAM, but not the stratified baseline hazard. We observed that the fitted baseline log cumulative hazard from HMC-Weibull is consistent with the overall shape of Cox-PG estimates (Figure \ref{fig:7}a). However, we observed violation of proportional hazards (linear effect) due to leukopenia in the Cox-PG estimates. If the effect was linear, there would be an equal-distance difference between the two baseline log cumulative hazards (Figure \ref{fig:7}a). Furthermore, the last death in the normal group was observed at day 704, whereas the leukopenia group recorded 87 deaths after day 704. This imbalanced distribution of event times suggests that stratified baseline hazards are more appropriate. We observed a decreased risk of mortality associated with the a low Townsend score reflecting affluent areas (Figure \ref{fig:7}b). The nonparametric effect through the affine combination of DR basis and penalized coefficients elucidates the nonlinear relationship between Townsend score and risk. The DR basis associated with quadratic relationship yielded a coefficient with a significant Bayesian p-value of 0.014 (Figure \ref{fig:7}c) indicating a nonlinear relationship. In both competing methods, the effect of age was significant, while sex effect was not (Table \ref{tab:1}).

\section{Discussion}

The link between Cox models and logistic regression is well known, with relative risks being equal to odds ratios in rare event settings. Because Cox models admits a local Poisson representation, we can reformulated Cox models as local negative binomial processes with a vague gamma mixture. These local negative binomial processes are in fact rare event logistic regressions with an appropriate offset term. This allows Bayesian computational and inferential tools of Gaussian inference through logistic regression to be deployed in survival settings. Now, multilevel Gaussian models can readily be adapted to Cox models. These techniques serve as a foundation for future work in Bayesian survival models. Cox-PG can be implemented in base \texttt{R} rather than working with a software framework such as \texttt{Stan} or \texttt{INLA}, making it more accessible to modelers. We provide our initial implementation of the Cox-PG algorithm in the supplement files. We leave other multilevel modeling extensions of Cox-PG and additional strategies for computational acceleration to future work.

\begin{figure}[H]
\centering
\includegraphics[width=5in]{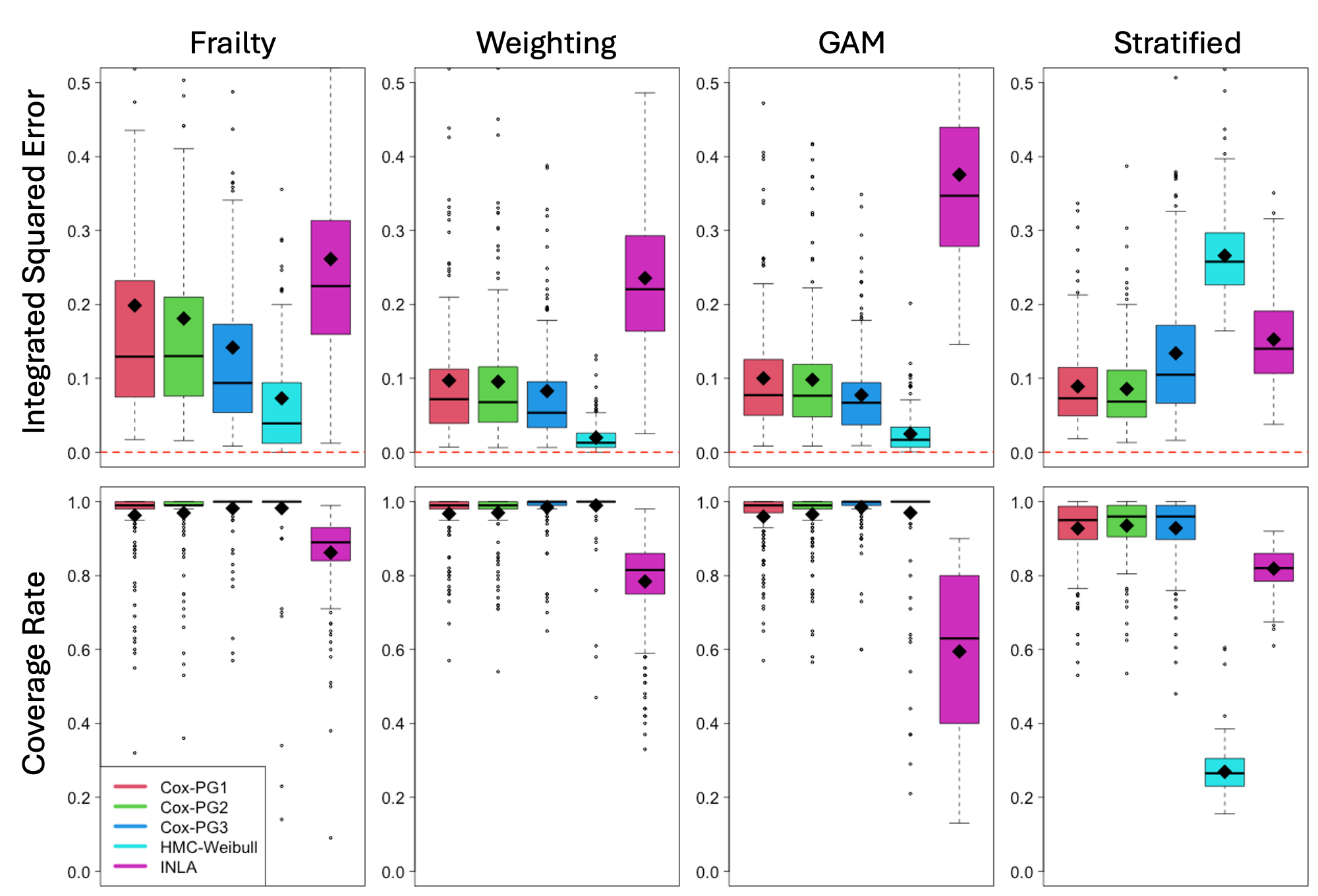}
{\linespread{1}\selectfont
\caption{Boxplot of integrated squared errors and integrated coverage rates for baseline log cumulative hazard. Means are denoted with diamond. }\label{fig:3}}
\end{figure}

\begin{table}[H]
\centering
\caption{Leukemia Regression Results for Competing Methods with 0.95 Credible Intervals}
\begin{tabular}{|l|cc|}
\hline
\textbf{Variable}      & \textbf{Cox-PG2}           & \textbf{HMC-Weibull}          \\
\hline
sex        & 0.0714 (-0.0680, 0.2187)          & 0.0573 (-0.0744,  0.1840)          \\
age        & 0.0314 (0.0272, 0.0358)       & 0.0316 (0.0275,  0.0356)       \\
\hline
\end{tabular}
\label{tab:1}
\end{table}

\bibliographystyle{agsm}
\bibliography{Bibliography-MM-MC}

\newpage
\section*{Supplement}
\subsection*{Simulation Studies: GAM results}

\begin{figure}[H]
\centering
\includegraphics[width=4in]{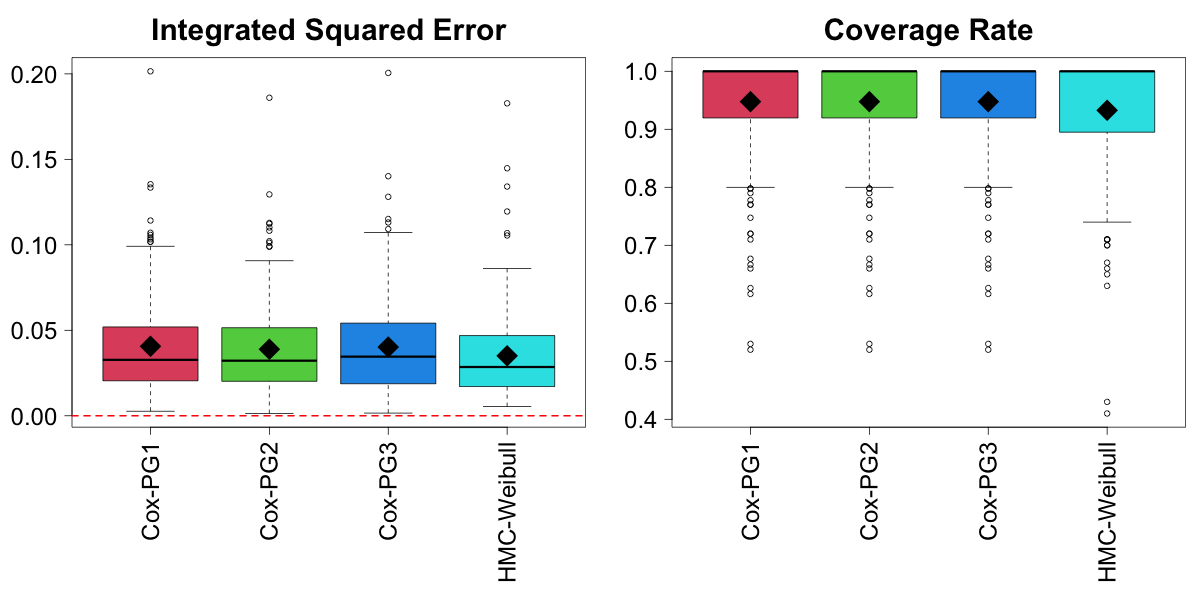}
{\linespread{1}\selectfont
\caption{Boxplot of integrated squared errors and integrated coverage rates for $\sin(x_3)$. Means are denoted with diamond. }\label{fig:3}}
\end{figure}

\subsection*{Appendix: Proofs}
\label{sec:appen}

In order to facilitate the proof Theorem 1 \& 2, we first present the following definitions and lemma.

\begin{definition}
\label{def1}
Truncated gamma density: $\tau \mid a_{0}, b_{0}, \tau_{0} \sim \mathrm{T G} \left(a_{0}, b_{0}, \tau_{0} \right)$
$$
\begin{array}{c}
\pi \left(\tau \mid a_{0}, b_{0}, \tau_{0}\right)=c_2 \left(\tau_{0}, a_{0}, b_{0}\right)^{-1} \tau^{a_{0}-1} \exp \left(-b_{0} \tau \right) \mathbb{I}\left(\tau \geq \tau_{0}\right)
\end{array}
$$
where $c_2 \left(\tau_{0}, a_{0}, b_{0}\right)=\int_{\tau_{0}}^{\infty} \tau^{a_{0}-1} \exp \left(-b_{0} \tau\right) d \tau$.
\end{definition}

\begin{definition}
\label{def2}
Loewner order of matrices $\mathbf{A} \succeq \mathbf{B}$: if $\mathbf{A}-\mathbf{B}$ is positive semi-definite or $\mathbf{u}^{\top}\mathbf{A}\mathbf{u} \geq \mathbf{u}^{\top}\mathbf{B}\mathbf{u}$ for all real $\mathbf{u}$ and the following relationships hold $| \mathbf{A}| \geq | \mathbf{B} |$ and $\mathbf{A}^{-1} \preceq \mathbf{B}^{-1}$.
\end{definition}

\begin{definition} Hyperbolic cosine through Laplace transform:
\label{def3}
$$
\begin{array}{c}
\int_{\mathbb{R}_{+}} \exp \left[-\frac{\left( \mathbf{m}_{i}^{\top} \boldsymbol{\eta}^\prime \right)^2+\left( \mathbf{m}_{i}^{\top} \boldsymbol{\eta} \right)^2}{2} \omega_i\right] \pi \left(\omega_i \mid y_i+\epsilon, 0 \right) d \omega_i = \left\{ \cosh \left[ \frac{\sqrt{\left( \mathbf{m}_{i}^{\top} \boldsymbol{\eta}^\prime \right)^2+\left( \mathbf{m}_{i}^{\top} \boldsymbol{\eta} \right)^2}}{2}\right] \right\}^{-(y_i+\epsilon)}
\end{array}
$$
\end{definition}

\begin{definition}
\label{def4}
Hyperbolic cosine inequalities: for $a, b \in \mathbb{R}, \cosh (a+b) \leq 2 \cosh (a) \cosh (b)$,  $\cosh (a) \geq 1$ and $\cosh (a) \leq \cosh (b)$ for $|a| \leq |b|$.
\end{definition}

\begin{lemma}
\label{lm1}
From Loewner ordering, we have 
$$
\begin{array}{rl}
( \boldsymbol{\eta}- \mathbf{Q}^{-1}\boldsymbol{\mu})^{\top} \mathbf{Q} ( \boldsymbol{\eta}- \mathbf{Q}^{-1} \boldsymbol{\mu}) 
\leq& 
\boldsymbol{\eta}^\top \mathbf{Q} \boldsymbol{\eta} -2 \boldsymbol{\eta}^\top \widetilde{\boldsymbol{\mu}} + 2 \boldsymbol{\eta}^\top \mathbf{M}^{\top} \boldsymbol{\Omega} \mathbf{1} \eta_\epsilon \\
& + 
\widetilde{\boldsymbol{\mu}}^{\top}  \mathbf{A}(\tau_0)^{-1} \widetilde{\boldsymbol{\mu}} + C_1 
\mathbf{1}^\top \boldsymbol{\Omega} \mathbf{1} + \eta_\epsilon  \mathbf{1}^\top \boldsymbol{\Omega} \mathbf{1} \eta_\epsilon 
\end{array}
$$
where $C_1 = \left\| 2 \eta_\epsilon \widetilde{\boldsymbol{\mu}}^{\top} \mathbf{A}(\tau_0)^{-1} \mathbf{M}^{\top} \right\|$ and $\mathbf{A}(\tau_0) = \boldsymbol{\Sigma}^{-1}_0 \oplus \tau_{0} \mathbf{I}_{M}$.
\end{lemma}

\begin{proof}Proof of Lemma \ref{lm1}.
Recall $\boldsymbol{\mu} = \widetilde{\boldsymbol{\mu}} - \mathbf{M}^{\top} \boldsymbol{\Omega} \mathbf{1} \eta_\epsilon$. We note that $\mathbf{Q} = \mathbf{M}^{\top} \boldsymbol{\Omega} \mathbf{M} + \mathbf{A}(\tau_{\mathcal{B}}) \succeq \mathbf{M}^{\top} \boldsymbol{\Omega} \mathbf{M}$ and $\mathbf{Q} \succeq \mathbf{A}(\tau_{\mathcal{B}}) \succeq \mathbf{A}( \tau_0 )$. Given that 
$$
\begin{array}{rl}
( \boldsymbol{\eta}- \mathbf{Q}^{-1}\boldsymbol{\mu})^{\top} \mathbf{Q} ( \boldsymbol{\eta}- \mathbf{Q}^{-1} \boldsymbol{\mu}) 
=& 
\boldsymbol{\eta}^\top \mathbf{Q} \boldsymbol{\eta}-2 \boldsymbol{\eta}^\top \boldsymbol{\mu} + \boldsymbol{\mu}^{\top} \mathbf{Q}^{-1} \boldsymbol{\mu} \\
=& 
\boldsymbol{\eta}^\top \mathbf{Q} \boldsymbol{\eta}-2 \boldsymbol{\eta}^\top \widetilde{\boldsymbol{\mu}} + 2 \boldsymbol{\eta}^\top \mathbf{M}^{\top} \boldsymbol{\Omega} \mathbf{1} \eta_\epsilon + \boldsymbol{\mu}^{\top} \mathbf{Q}^{-1} \boldsymbol{\mu} .
\end{array}
$$
We apply Loewner ordering to the projection matrix 
$$
\begin{array}{c}
\mathbf{Q}^{-1} \preceq (\mathbf{M}^{\top} \boldsymbol{\Omega} \mathbf{M})^{-1}\\
\boldsymbol{\Omega}^{1/2} \mathbf{M} (\mathbf{M}^{\top} \boldsymbol{\Omega} \mathbf{M})^{-1} \mathbf{M}^{\top} \boldsymbol{\Omega}^{1/2} \preceq \mathbf{I},
\end{array}
$$ triangle and Cauchy–Schwarz inequality next,
$$
\begin{array}{rl}
\boldsymbol{\mu}^{\top} \mathbf{Q}^{-1} \boldsymbol{\mu} 
=& 
\widetilde{\boldsymbol{\mu}}^{\top} \mathbf{Q}^{-1} \widetilde{\boldsymbol{\mu}} - 2 \widetilde{\boldsymbol{\mu}}^{\top} \mathbf{Q}^{-1} \mathbf{M}^{\top} \boldsymbol{\Omega} \mathbf{1} \eta_\epsilon + \eta_\epsilon  \mathbf{1}^\top \boldsymbol{\Omega} \mathbf{M} \mathbf{Q}^{-1} \mathbf{M}^{\top} \boldsymbol{\Omega} \mathbf{1} \eta_\epsilon \\
\leq&
\widetilde{\boldsymbol{\mu}}^{\top} \mathbf{Q}^{-1} \widetilde{\boldsymbol{\mu}} - 2 \widetilde{\boldsymbol{\mu}}^{\top} \mathbf{Q}^{-1} \mathbf{M}^{\top} \boldsymbol{\Omega} \mathbf{1} \eta_\epsilon + \eta_\epsilon  \mathbf{1}^\top \boldsymbol{\Omega} \mathbf{1} \eta_\epsilon \\
\leq&
\widetilde{\boldsymbol{\mu}}^{\top} \mathbf{Q}^{-1} \widetilde{\boldsymbol{\mu}} + \left| 2 \eta_\epsilon \widetilde{\boldsymbol{\mu}}^{\top} \mathbf{Q}^{-1} \mathbf{M}^{\top} \boldsymbol{\Omega} \mathbf{1} \right| + \eta_\epsilon  \mathbf{1}^\top \boldsymbol{\Omega} \mathbf{1} \eta_\epsilon \\
\leq&
\widetilde{\boldsymbol{\mu}}^{\top} \mathbf{Q}^{-1} \widetilde{\boldsymbol{\mu}} + \left\| 2 \eta_\epsilon \widetilde{\boldsymbol{\mu}}^{\top} \mathbf{Q}^{-1} \mathbf{M}^{\top} \right\| \left\| \boldsymbol{\Omega} \mathbf{1} \right\| + \eta_\epsilon  \mathbf{1}^\top \boldsymbol{\Omega} \mathbf{1} \eta_\epsilon \\
\leq&
\widetilde{\boldsymbol{\mu}}^{\top}  \mathbf{A}(\tau_0)^{-1} \widetilde{\boldsymbol{\mu}} + C_1 
\mathbf{1}^\top \boldsymbol{\Omega} \mathbf{1} + \eta_\epsilon  \mathbf{1}^\top \boldsymbol{\Omega} \mathbf{1} \eta_\epsilon 
\end{array}
$$
where $\left\| \boldsymbol{\Omega} \mathbf{1} \right\| = \sqrt{ \sum_i \omega_i^2 } \leq \sum_i \omega_i = \mathbf{1}^\top \boldsymbol{\Omega} \mathbf{1}$ and $C_1 = \left\| 2 \eta_\epsilon \widetilde{\boldsymbol{\mu}}^{\top} \mathbf{A}(\tau_0)^{-1} \mathbf{M}^{\top} \right\| \geq \left\| 2 \eta_\epsilon \widetilde{\boldsymbol{\mu}}^{\top} \mathbf{Q}^{-1} \mathbf{M}^{\top} \right\|$.
\end{proof}

\begin{proof}Proof of Theorem 1.
First we use the truncated gamma density to replace the gamma conditional distribution 
$$
\begin{array}{c}
(\tau_\mathcal{B} \mid -) \sim \mathrm{T G} \left(a_{0}+\frac{M}{2}, b_{0}+\frac{ \mathbf{u}^{\top}_\mathcal{B} \mathbf{u}_\mathcal{B}}{2}, \tau_{0}\right).
\end{array}
$$
We note that $\mathbf{Q} = \mathbf{M}^{\top} \boldsymbol{\Omega} \mathbf{M} + \mathbf{A}(\tau_{\mathcal{B}}) \succeq \mathbf{M}^{\top} \boldsymbol{\Omega} \mathbf{M} + \mathbf{A}(\tau_0) \succeq \mathbf{A}(\tau_0)$ and 
$$ 
\begin{array}{rl}
c_1 ( \mathbf{Q}^{-1} \boldsymbol{\mu}, \mathbf{Q}^{-1}, \mathcal{C}_0 \bigcap \mathcal{C}_v ) &= \oint_{ \boldsymbol{\eta} \in \mathcal{C}_0 \bigcap \mathcal{C}_v } \exp \left\{-\frac{1}{2}( \boldsymbol{\eta}- \mathbf{Q}^{-1}\boldsymbol{\mu})^{\top} \mathbf{Q} ( \boldsymbol{\eta}- \mathbf{Q}^{-1}\boldsymbol{\mu})\right\} d \boldsymbol{\eta} \\
&\leq \int \exp \left\{-\frac{1}{2}( \boldsymbol{\eta}- \mathbf{Q}^{-1}\boldsymbol{\mu})^{\top} \mathbf{Q} ( \boldsymbol{\eta}- \mathbf{Q}^{-1}\boldsymbol{\mu})\right\} d \boldsymbol{\eta} \\
&= (2 \pi)^{(P+M)/2} | \mathbf{Q} |^{-1/2} \\
c_1 ( \mathbf{Q}^{-1} \boldsymbol{\mu}, \mathbf{Q}^{-1}, \mathcal{C}_0 \bigcap \mathcal{C}_v )^{-1} &\geq (2 \pi)^{-(P+M)/2} | \mathbf{Q} |^{1/2}
\end{array}
$$
with $| \mathbf{Q} | \geq | \mathbf{A}(\tau_0) |$. In addition, $\mathbb{I}( \boldsymbol{\eta} \in \mathcal{C}_v \bigcap \mathcal{C}_0 ) = \mathbb{I}( \boldsymbol{\eta} \in \mathcal{C}_0 )\mathbb{I}( \boldsymbol{\eta} \in \mathcal{C}_v )$. After applying Lemma \ref{lm1}, we have the inequality, 

$$
\begin{array}{rl}
\pi(\boldsymbol{\eta} \mid \boldsymbol{\omega}, \tau_{\mathcal{B}}, \mathbf{v}, \mathbf{y}) 
=& c_1 ( \mathbf{Q}^{-1} \boldsymbol{\mu}, \mathbf{Q}^{-1}, \mathcal{C}_0 \bigcap \mathcal{C}_v )^{-1} 
\mathbb{I}( \boldsymbol{\eta} \in \mathcal{C}_v \bigcap \mathcal{C}_0 )\\
& \times \exp \left[-\frac{1}{2}\left(\boldsymbol{\eta}-\mathbf{Q}^{-1} \boldsymbol{\mu}\right)^\top \mathbf{Q}\left(\boldsymbol{\eta}-\mathbf{Q}^{-1} \boldsymbol{\mu}\right)\right] \\
\geq &(2 \pi)^{-\frac{P+M}{2}}\left|\mathbf{A}\left(\tau_0\right)\right|^{1 / 2} \mathbb{I}( \boldsymbol{\eta} \in \mathcal{C}_0 )\mathbb{I}( \boldsymbol{\eta} \in \mathcal{C}_v ) \\
& \times \exp \left[ -\frac{1}{2} \left( \boldsymbol{\eta}^\top \mathbf{Q} \boldsymbol{\eta} + 2 \boldsymbol{\eta}^\top \mathbf{M}^{\top} \boldsymbol{\Omega} \mathbf{1} \eta_\epsilon + \eta_\epsilon  \mathbf{1}^\top \boldsymbol{\Omega} \mathbf{1} \eta_\epsilon \right) \right] \\
& \times \exp \left[ -\frac{1}{2} \left\{ -2 \boldsymbol{\eta}^\top \widetilde{\boldsymbol{\mu}} + 
\widetilde{\boldsymbol{\mu}}^{\top}  \mathbf{A}(\tau_0)^{-1} \widetilde{\boldsymbol{\mu}} + C_1 
\mathbf{1}^\top \boldsymbol{\Omega} \mathbf{1} \right\} \right] \\
=& (2 \pi)^{-\frac{P+M}{2}}\left|\mathbf{A}\left(\tau_0\right)\right|^{1 / 2} \mathbb{I}( \boldsymbol{\eta} \in \mathcal{C}_0 )\mathbb{I}( \boldsymbol{\eta} \in \mathcal{C}_v ) \\
& \times \exp \left[-\frac{1}{2} \boldsymbol{\beta}^\top \boldsymbol{\Sigma}^{-1}_0 \boldsymbol{\beta}-\frac{1}{2} \tau_{\mathcal{B}} \mathbf{u}_{\mathcal{B}} ^\top \mathbf{u}_{\mathcal{B}} \right]  \exp \left[-\frac{1}{2} \sum_{i=1}^N \omega_i\left(\mathbf{m}_i^\top \boldsymbol{\eta}\right)^2\right] \\
& \times \exp \left[ -\frac{1}{2} \left( 2 \boldsymbol{\eta}^\top \mathbf{M}^{\top} \boldsymbol{\Omega} \mathbf{1} \eta_\epsilon + \eta_\epsilon \mathbf{1}^\top \boldsymbol{\Omega} \mathbf{1} \eta_\epsilon + C_1 
\mathbf{1}^\top \boldsymbol{\Omega} \mathbf{1} \right) \right] \\
& \times \exp \left[ -\frac{1}{2} \left\{ -2 \boldsymbol{\eta}^\top \widetilde{\boldsymbol{\mu}} + 
\widetilde{\boldsymbol{\mu}}^{\top}  \mathbf{A}(\tau_0)^{-1} \widetilde{\boldsymbol{\mu}} \right\} \right].
\end{array}
$$

Note that we expand the quadratic form of $\boldsymbol{\eta}^\top \mathbf{Q} \boldsymbol{\eta} = \boldsymbol{\eta}^\top \left\{ \mathbf{M}^{\top} \boldsymbol{\Omega} \mathbf{M} + \mathbf{A}(\tau_{\mathcal{B}}) \right\} \boldsymbol{\eta}$. With some algebra to combine $\sum_{i=1}^N \omega_i\left(\mathbf{m}_i^\top \boldsymbol{\eta}\right)^2$, $2 \boldsymbol{\eta}^\top \mathbf{M}^{\top} \boldsymbol{\Omega} \mathbf{1} \eta_\epsilon$ and $\eta_\epsilon  \mathbf{1}^\top \boldsymbol{\Omega} \mathbf{1} \eta_\epsilon$ to get $\sum_{i=1}^N \omega_i\left(\mathbf{m}_i^\top \boldsymbol{\eta}  + {\eta}_\epsilon \right)^2$. We have 
$$
\begin{array}{rl}
\pi(\boldsymbol{\eta} \mid \boldsymbol{\omega}, \tau_{\mathcal{B}}, \mathbf{v}, \mathbf{y}) 
\geq&
(2 \pi)^{-\frac{P+M}{2}}\left|\mathbf{A}\left(\tau_0\right)\right|^{1 / 2} \mathbb{I}( \boldsymbol{\eta} \in \mathcal{C}_0 )\mathbb{I}( \boldsymbol{\eta} \in \mathcal{C}_v ) \\
& \times \exp \left[-\frac{1}{2} \boldsymbol{\beta}^\top \boldsymbol{\Sigma}^{-1}_0 \boldsymbol{\beta}-\frac{1}{2} \tau_{\mathcal{B}} \mathbf{u}_{\mathcal{B}} ^\top \mathbf{u}_{\mathcal{B}} \right] \exp \left[-\frac{1}{2} \sum_{i=1}^N \omega_i\left(\mathbf{m}_i^\top \boldsymbol{\eta}  + {\eta}_\epsilon \right)^2\right] \\
& \times \exp \left[ -\frac{1}{2} \left\{ -2 \boldsymbol{\eta}^\top \widetilde{\boldsymbol{\mu}} + 
\widetilde{\boldsymbol{\mu}}^{\top}  \mathbf{A}(\tau_0)^{-1} \widetilde{\boldsymbol{\mu}} + C_1 
\mathbf{1}^\top \boldsymbol{\Omega} \mathbf{1} \right\} \right] \\
=&
(2 \pi)^{-\frac{P+M}{2}}\left|\mathbf{A}\left(\tau_0\right)\right|^{1 / 2} \mathbb{I}( \boldsymbol{\eta} \in \mathcal{C}_0 )\mathbb{I}( \boldsymbol{\eta} \in \mathcal{C}_v ) \\
& \times \exp \left[-\frac{1}{2} \boldsymbol{\beta}^\top \boldsymbol{\Sigma}^{-1}_0 \boldsymbol{\beta}-\frac{1}{2} \tau_{\mathcal{B}} \mathbf{u}_{\mathcal{B}} ^\top \mathbf{u}_{\mathcal{B}} \right] \exp \left[-\frac{1}{2} \sum_{i=1}^N \omega_i \psi_i^2\right] \\
& \times \exp \left[ -\frac{1}{2} \left\{ -2 \boldsymbol{\eta}^\top \widetilde{\boldsymbol{\mu}} + 
\widetilde{\boldsymbol{\mu}}^{\top}  \mathbf{A}(\tau_0)^{-1} \widetilde{\boldsymbol{\mu}} + C_1 
\mathbf{1}^\top \boldsymbol{\Omega} \mathbf{1} \right\} \right]
\end{array}
$$
with $\psi_i = \mathbf{m}_{i}^{\top} \boldsymbol{\eta} + \eta_\epsilon$. 

Replacing our gamma Markov chain transitions with truncated gamma $\pi \left(\tau_\mathcal{B} \mid a_{0}, b_{0}, \tau_{0}\right)$, following Theorem 1 of \citealt{wang2018analysis}, we now bound 
$$
\begin{array}{rl}
g \left( \boldsymbol{\eta} \mid \mathbf{v} \right) =& 
\int_{\mathbb{R}_{+}} \int_{\mathbb{R}_{+}^{N}} \pi( \boldsymbol{\eta} \mid \boldsymbol{\omega}, \tau_{\mathcal{B}}, \mathbf{v}, \mathbf{y}) \pi\left(\boldsymbol{\omega}, \tau_{\mathcal{B}} \mid \boldsymbol{\eta}^{\prime}, \mathbf{y}\right) d \boldsymbol{\omega} d \tau_{\mathcal{B}} \\
=& \int_{\mathbb{R}_{+}} \int_{\mathbb{R}_{+}^{N}} \pi( \boldsymbol{\eta} \mid \boldsymbol{\omega}, \tau_{\mathcal{B}}, \mathbf{v}, \mathbf{y}) \pi \left(\boldsymbol{\omega} \mid \boldsymbol{\eta}^{\prime}, \mathbf{y}\right) \pi \left( \tau_{\mathcal{B}} \mid \boldsymbol{\eta}^{\prime}, \mathbf{y}\right) d \boldsymbol{\omega} d \tau_{\mathcal{B}}. 
\end{array}
$$
From Theorem 1 of \citealt{wang2018analysis}, under condition (C\ref{c1}), we have 
$$
\begin{array}{rl}
\int_{\mathbb{R}_{+}} \pi( \boldsymbol{\eta} \mid \boldsymbol{\omega}, \tau_{\mathcal{B}}, \mathbf{v}, \mathbf{y}) \pi \left( \tau_{\mathcal{B}} \mid \boldsymbol{\eta}^{\prime}, \mathbf{y}\right) d \tau_{\mathcal{B}} \geq&
(2 \pi)^{-\frac{P+M}{2}}\left|\mathbf{A}\left(\tau_0\right)\right|^{1 / 2} \mathbb{I}( \boldsymbol{\eta} \in \mathcal{C}_0 )\mathbb{I}( \boldsymbol{\eta} \in \mathcal{C}_v ) \\
& \times \exp \left[-\frac{1}{2} \boldsymbol{\beta}^\top \boldsymbol{\Sigma}^{-1}_0 \boldsymbol{\beta} \right] \exp \left[-\frac{1}{2} \sum_{i=1}^N \omega_i \psi_i^2\right] \\
& \times \exp \left[ -\frac{1}{2} \left\{ -2 \boldsymbol{\eta}^\top \widetilde{\boldsymbol{\mu}} + 
\widetilde{\boldsymbol{\mu}}^{\top}  \mathbf{A}(\tau_0)^{-1} \widetilde{\boldsymbol{\mu}} + C_1 
\mathbf{1}^\top \boldsymbol{\Omega} \mathbf{1} \right\} \right] \\
& \times \exp \left[ -\frac{1}{2} \tau_{0} \mathbf{u}_{\mathcal{B}} ^\top \mathbf{u}_{\mathcal{B}} \right] \left(\frac{b_0}{b_0+\mathbf{u}_{\mathcal{B}}^\top \mathbf{u}_{\mathcal{B}} / 2}\right)^{a_0+M / 2} .
\end{array}
$$
The next step of the proof follows the spirit of \citealt{wang2018analysis}, but it is less straight forward so we write these steps in detail. Note that the conditional distribution, $\pi \left(\boldsymbol{\omega} \mid \boldsymbol{\eta}^{\prime}, \mathbf{y}\right)$ is given as $\omega_i \mid \boldsymbol{\eta}^\prime, \mathbf{y} \sim \mathrm{PG}\left(y_i+\epsilon, \psi_i^\prime \right)$ with $\psi^\prime_i = \mathbf{m}_{i}^{\top} \boldsymbol{\eta}^\prime + \eta_\epsilon$ and $C_1 \mathbf{1}^\top \boldsymbol{\Omega} \mathbf{1} = \sum_{i=1}^N C_1 \omega_i$. After multiplying by $\pi \left(\boldsymbol{\omega} \mid \boldsymbol{\eta}^{\prime}, \mathbf{y}\right)$, using Definition \ref{def3} \& \ref{def4}, we carry out the integral
$$
\begin{array}{rl}
\int_{\mathbb{R}_{+}} \exp\left( -\frac{1}{2} \omega_i \psi_i^2 - \frac{1}{2} \omega_i C_1 \right) \pi(\omega_i \mid y_i + \epsilon, \psi_i) d \omega_i
= &
\int_{\mathbb{R}_{+}} \exp\left( -\frac{1}{2} \omega_i \psi_i^2 - \frac{1}{2} \omega_i C_1 \right) \\
& \times 
\cosh^{y_i+\epsilon} ( |\psi^{\prime}_i / 2| ) \exp \left[ -\frac{1}{2} \omega_i (\psi_i^{\prime})^2 \right] \\
& \times \pi (\omega_i \mid y_i+\epsilon, 0) d \omega_i \\
= & \left\{ \cosh \left[ \frac{\sqrt{\left( \psi_i^\prime \right)^2+\left( \psi_i \right)^2+C_1}}{2}\right] \right\}^{-(y_i+\epsilon)} \\
& \times \cosh^{y_i+\epsilon} ( |\psi^{\prime}_i / 2| ) \\
\geq &
\left\{ \cosh \left[ \frac{ \left| \psi_i^\prime \right| + \left| \psi_i \right| + \sqrt{C_1} }{2} \right] \right\}^{-(y_i+\epsilon)} \\
& \times \cosh^{y_i+\epsilon} ( |\psi^{\prime}_i / 2| ) \\
\geq &
\Bigg\{ 4 \cosh \left(  \left| \psi_i^\prime \right| / 2 \right) \cosh \left( \left| \psi_i \right| / 2 \right) \\
& \times \cosh \left( \sqrt{C_1} /2  \right) \Bigg\}^{-(y_i+\epsilon)}  \cosh^{y_i+\epsilon} ( |\psi^{\prime}_i / 2| ).
\end{array} 
$$
We have 
$$
\begin{array}{c}
\int_{\mathbb{R}_{+}} \exp\left( -\frac{1}{2} \omega_i \psi_i^2 - \frac{1}{2}C_1 \omega_i \right) \pi(\omega_i \mid y_i + \epsilon, \psi_i) d \omega_i \geq \Bigg\{ 4 \cosh \left( \left| \psi_i \right| / 2 \right) \cosh \left( \sqrt{C_1} /2  \right) \Bigg\}^{-(y_i+\epsilon)}.
\end{array} 
$$
Note that 
$$
\begin{array}{rl}
\cosh \left( \left| \psi_i \right| / 2 \right) \geq \left[\exp \left(\frac{\left| \psi_i \right|}{2}\right)\right]^{-(y_i+\epsilon)} \geq \left[\exp \left(\frac{ \psi_i^2 + 1}{4}\right)\right]^{-(y_i+\epsilon)}= e^{- (y_i + \epsilon) / 4} \exp \left[-\frac{y_i + \epsilon}{4} \psi_i^2 \right]
\end{array} 
$$
with quadratic expansion $\sum_{i=1}^N (y_i + \epsilon) \psi_i^2 = \left[ \boldsymbol{\eta}, \eta_\epsilon \right]^\top \left[ \mathbf{M}, \mathbf{1} \right]^\top \boldsymbol{\Lambda} \left[ \mathbf{M}, \mathbf{1} \right] \left[ \boldsymbol{\eta}, \eta_\epsilon \right] $ where  $\boldsymbol{\Lambda}$ is the $N \times N$ diagonal matrix with $i$th diagonal element $y_i+\epsilon$. We can simplify the expansion as  $\sum_{i=1}^N (y_i + \epsilon) \psi_i^2 = \boldsymbol{\eta}^\top \mathbf{M}^\top \boldsymbol{\Lambda} \mathbf{M} \boldsymbol{\eta} + 2 \boldsymbol{\eta}^\top \mathbf{M}^{\top} \boldsymbol{\Lambda} \mathbf{1} \eta_\epsilon + \eta_\epsilon  \mathbf{1}^\top \boldsymbol{\Lambda} \mathbf{1} \eta_\epsilon$. 
$$
\begin{array}{rl}
g \left( \boldsymbol{\eta} \mid \mathbf{v} \right) =& 
\int_{\mathbb{R}_{+}} \int_{\mathbb{R}_{+}^{N}} \pi( \boldsymbol{\eta} \mid \boldsymbol{\omega}, \tau_{\mathcal{B}}, \mathbf{v}, \mathbf{y}) \pi\left(\boldsymbol{\omega}, \tau_{\mathcal{B}} \mid \boldsymbol{\eta}^{\prime}, \mathbf{y}\right) d \boldsymbol{\omega} d \tau_{\mathcal{B}} \\
\geq & (2 \pi)^{-\frac{P+M}{2}}\left|\mathbf{A}\left(\tau_0\right)\right|^{1 / 2} \left[ 4 \cosh( \sqrt{C_1} /2) \right]^{-N_\epsilon} e^{-\frac{N_\epsilon}{4}} \mathbb{I}( \boldsymbol{\eta} \in \mathcal{C}_0 )\mathbb{I}( \boldsymbol{\eta} \in \mathcal{C}_v ) \\
& \times \exp \left[-\frac{1}{2} \boldsymbol{\beta}^\top \boldsymbol{\Sigma}^{-1}_0 \boldsymbol{\beta}+\boldsymbol{\eta}^\top \widetilde{\boldsymbol{\mu}}-\frac{1}{2} \widetilde{\boldsymbol{\mu}}^\top \mathbf{A}\left(\tau_0\right)^{-1} \widetilde{\boldsymbol{\mu}} \right] \exp \left(-\tau_0 \mathbf{u}_{\mathcal{B}}^\top \mathbf{u}_{\mathcal{B}} / 2\right) \\
& \times \exp \left[-\frac{1}{4} \boldsymbol{\eta}^\top \mathbf{M}^\top \boldsymbol{\Lambda} \mathbf{M} \boldsymbol{\eta} - \frac{1}{2} \boldsymbol{\eta}^\top \mathbf{M}^{\top} \boldsymbol{\Lambda} \mathbf{1} \eta_\epsilon - \frac{1}{4} \eta_\epsilon  \mathbf{1}^\top \boldsymbol{\Lambda} \mathbf{1} \eta_\epsilon \right] \left(\frac{b_0}{b_0+\mathbf{u}_{\mathcal{B}}^\top \mathbf{u}_{\mathcal{B}} / 2}\right)^{a_0+M / 2} 
\end{array}
$$
where $N_\epsilon = \sum_{i=1}^N y_i+\epsilon$. Let $\overline{\boldsymbol{\mu}} = - \mathbf{M}^{\top} \boldsymbol{\Lambda} \mathbf{1} \eta_\epsilon / 2$, $\mathbf{H} = \mathbf{M}^\top \boldsymbol{\Lambda} \mathbf{M}/2$ and we get
$$
\begin{array}{rl}
\exp \left[-\frac{1}{4} \boldsymbol{\eta}^\top \mathbf{M}^\top \boldsymbol{\Lambda} \mathbf{M} \boldsymbol{\eta} - \frac{1}{2} \boldsymbol{\eta}^\top \mathbf{M}^{\top} \boldsymbol{\Lambda} \mathbf{1} \eta_\epsilon \right] =& \exp \left[-\frac{1}{2} \left( \boldsymbol{\eta} - \mathbf{H}^{-1} \overline{\boldsymbol{\mu}} \right)^\top \mathbf{H} \left( \boldsymbol{\eta} - \mathbf{H}^{-1} \overline{\boldsymbol{\mu}} \right) \right] \\
& \times \exp \left( \frac{1}{2} \overline{\boldsymbol{\mu}}^\top \mathbf{H}^{-1} \overline{\boldsymbol{\mu}} \right) .
\end{array}
$$
Note that $\overline{\boldsymbol{\mu}}^\top \mathbf{H}^{-1} \overline{\boldsymbol{\mu}} \leq \eta_\epsilon  \mathbf{1}^\top \boldsymbol{\Lambda} \mathbf{1} \eta_\epsilon / 2$ by Loewner ordering (Definition \ref{def2}).
Let $\boldsymbol{\mu}_{\Lambda} = \widetilde{\boldsymbol{\mu}} + \overline{\boldsymbol{\mu}}$ and 
$$
\begin{array}{c}
R_1 = \left[ 4 \cosh( \sqrt{C_1} /2) \right]^{-N_\epsilon} e^{-\frac{N_\epsilon}{4}} \exp \left( - \frac{1}{2} \widetilde{\boldsymbol{\mu}}^\top \mathbf{A}\left(\tau_0\right)^{-1} \widetilde{\boldsymbol{\mu}} - \frac{1}{4} \eta_\epsilon  \mathbf{1}^\top \boldsymbol{\Lambda} \mathbf{1} \eta_\epsilon \right) < 1
\end{array}
$$
(see Definition \ref{def4} for $\cosh$ relationship).
We can write the bounds of $g \left( \boldsymbol{\eta} \mid \mathbf{v} \right)$ as a Gaussian kernel,
$$
\begin{array}{rl}
g \left( \boldsymbol{\eta} \mid \mathbf{v} \right) 
\geq & 
(2 \pi)^{-\frac{P+M}{2}}\left|\mathbf{A}\left(\tau_0\right)\right|^{1 / 2} \left[ 4 \cosh( \sqrt{C_1} /2) \right]^{-N_\epsilon} e^{-\frac{N_\epsilon}{4}} \left(\frac{b_0}{b_0+\mathbf{u}_{\mathcal{B}}^\top \mathbf{u}_{\mathcal{B}} / 2}\right)^{a_0+M / 2} \\
& \times \exp \left[ -\frac{1}{2} \left\{ \boldsymbol{\eta} - \mathbf{A}(\tau_0)^{-1} \widetilde{\boldsymbol{\mu}} \right\}^{\top} \mathbf{A}(\tau_0) \left\{ \boldsymbol{\eta} - \mathbf{A}(\tau_0)^{-1} \widetilde{\boldsymbol{\mu}} \right\} \right] \\
& \times \exp \left[-\frac{1}{2} \left( \boldsymbol{\eta} - \mathbf{H}^{-1} \overline{\boldsymbol{\mu}} \right)^\top \mathbf{H} \left( \boldsymbol{\eta} - \mathbf{H}^{-1} \overline{\boldsymbol{\mu}} \right) \right] \\
& \times \exp \left( \frac{1}{2} \overline{\boldsymbol{\mu}}^\top \mathbf{H}^{-1} \overline{\boldsymbol{\mu}} - \frac{1}{4} \eta_\epsilon  \mathbf{1}^\top \boldsymbol{\Lambda} \mathbf{1} \eta_\epsilon \right) \mathbb{I}( \boldsymbol{\eta} \in \mathcal{C}_0 )\mathbb{I}( \boldsymbol{\eta} \in \mathcal{C}_v ) \\
= & 
(2 \pi)^{-\frac{P+M}{2}}\left|\mathbf{A}\left(\tau_0\right)\right|^{1 / 2} \left[ 4 \cosh( \sqrt{C_1} /2) \right]^{-N_\epsilon} e^{-\frac{N_\epsilon}{4}} \left(\frac{b_0}{b_0+\mathbf{u}_{\mathcal{B}}^\top \mathbf{u}_{\mathcal{B}} / 2}\right)^{a_0+M / 2} \\
& \times \exp \left( -\frac{1}{2} \left\{ \boldsymbol{\eta} - \mathbf{S}^{-1} \boldsymbol{\mu}_{\Lambda} \right\}^{\top} \mathbf{S} \left\{ \boldsymbol{\eta} - \mathbf{S}^{-1} \boldsymbol{\mu}_{\Lambda} \right\} \right) \\
& \times \exp \left[ \frac{1}{2} \left\{ \boldsymbol{\mu}_{\Lambda}^\top \mathbf{S}^{-1} \boldsymbol{\mu}_{\Lambda} - \widetilde{\boldsymbol{\mu}}^\top \mathbf{A}\left(\tau_0\right)^{-1} \widetilde{\boldsymbol{\mu}} - \overline{\boldsymbol{\mu}}^\top \mathbf{H}^{-1} \overline{\boldsymbol{\mu}} \right\} \right] \\
& \times \exp \left( \frac{1}{2} \overline{\boldsymbol{\mu}}^\top \mathbf{H}^{-1} \overline{\boldsymbol{\mu}} - \frac{1}{4} \eta_\epsilon  \mathbf{1}^\top \boldsymbol{\Lambda} \mathbf{1} \eta_\epsilon \right) \mathbb{I}( \boldsymbol{\eta} \in \mathcal{C}_0 )\mathbb{I}( \boldsymbol{\eta} \in \mathcal{C}_v ) \\
\geq & 
(2 \pi)^{-\frac{P+M}{2}}\left|\mathbf{A}\left(\tau_0\right)\right|^{1 / 2} \left[ 4 \cosh( \sqrt{C_1} /2) \right]^{-N_\epsilon} e^{-\frac{N_\epsilon}{4}} \left(\frac{b_0}{b_0+\mathbf{u}_{\mathcal{B}}^\top \mathbf{u}_{\mathcal{B}} / 2}\right)^{a_0+M / 2} \\
& \times \exp \left( -\frac{1}{2} \left\{ \boldsymbol{\eta} - \mathbf{S}^{-1} \boldsymbol{\mu}_{\Lambda} \right\}^{\top} \mathbf{S} \left\{ \boldsymbol{\eta} - \mathbf{S}^{-1} \boldsymbol{\mu}_{\Lambda} \right\} \right) \\
& \times \exp \left[ - \frac{1}{2} \widetilde{\boldsymbol{\mu}}^\top \mathbf{A}\left(\tau_0\right)^{-1} \widetilde{\boldsymbol{\mu}} - \frac{1}{4} \eta_\epsilon  \mathbf{1}^\top \boldsymbol{\Lambda} \mathbf{1} \eta_\epsilon \right]
\mathbb{I}( \boldsymbol{\eta} \in \mathcal{C}_0 )\mathbb{I}( \boldsymbol{\eta} \in \mathcal{C}_v ) \\
= &
(2 \pi)^{-\frac{P+M}{2}}\left|\mathbf{A}\left(\tau_0\right)\right|^{1 / 2} R_1 \left(\frac{b_0}{b_0+\mathbf{u}_{\mathcal{B}}^\top \mathbf{u}_{\mathcal{B}} / 2}\right)^{a_0+M / 2} \mathbb{I}( \boldsymbol{\eta} \in \mathcal{C}_0 )\mathbb{I}( \boldsymbol{\eta} \in \mathcal{C}_v )  \\
& \times \exp \left( -\frac{1}{2} \left\{ \boldsymbol{\eta} - \mathbf{S}^{-1} \boldsymbol{\mu}_{\Lambda} \right\}^{\top} \mathbf{S} \left\{ \boldsymbol{\eta} - \mathbf{S}^{-1} \boldsymbol{\mu}_{\Lambda} \right\} \right) 
\end{array}
$$
where $\mathbf{S}= \mathbf{M}^\top \boldsymbol{\Lambda} \mathbf{M} / 2 + \mathbf{A}(\tau_0) = \mathbf{H} + \mathbf{A}(\tau_0) \succeq \mathbf{A}\left(\tau_0\right)$. 

Note that $\mathbb{I}( \boldsymbol{\eta} \in \mathcal{C}_v ) = \prod_{j=1}^J \mathbb{I} ( v_j \leq {u}_{\alpha,j} )$ and using (C\ref{c2}) with Theorem 7 from \citealt{mira2002efficiency}, we get
$$
\begin{array}{rl}
k\left( \boldsymbol{\eta} \mid \boldsymbol{\eta}^{\prime}\right)
= &
\int_{\mathbb{R}_{+}^{J}} \int_{\mathbb{R}_{+}} \int_{\mathbb{R}_{+}^{N}} \pi( \boldsymbol{\eta} \mid \boldsymbol{\omega}, \tau_{\mathcal{B}}, \mathbf{v}, \mathbf{y}) \pi\left(\boldsymbol{\omega}, \tau_{\mathcal{B}}, \mathbf{v} \mid \boldsymbol{\eta}^{\prime}, \mathbf{y}\right) d \boldsymbol{\omega} d \tau_{\mathcal{B}} d \mathbf{v} \\
\geq & (2 \pi)^{-\frac{P+M}{2}}\left|\mathbf{A}\left(\tau_0\right)\right|^{1 / 2} R_1 \left(\frac{b_0}{b_0+\mathbf{u}_{\mathcal{B}}^\top \mathbf{u}_{\mathcal{B}} / 2}\right)^{a_0+M / 2} \mathbb{I}( \boldsymbol{\eta} \in \mathcal{C}_0 ) \\
& \times \exp \left( -\frac{1}{2} \left\{ \boldsymbol{\eta} - \mathbf{S}^{-1} \boldsymbol{\mu}_{\Lambda} \right\}^{\top} \mathbf{S} \left\{ \boldsymbol{\eta} - \mathbf{S}^{-1} \boldsymbol{\mu}_{\Lambda} \right\} \right) \\
& \times \int_{\mathbb{R}_{+}^{J}} \mathbb{I}( \boldsymbol{\eta} \in \mathcal{C}_v ) \pi( \mathbf{v} \mid \boldsymbol{\eta}^{\prime} )  d \mathbf{v} 
\end{array}
$$
where 
$$
\begin{array}{rl}
\int_{\mathbb{R}_{+}^{J}} \mathbb{I}( \boldsymbol{\eta} \in \mathcal{C}_v ) \pi( \mathbf{v} \mid \boldsymbol{\eta}^{\prime} )  d \mathbf{v} 
=&
\int_{\mathbb{R}_{+}^{J}} \prod_{j=1}^J
\frac{ n_{\alpha,j} v_j^{n_{\alpha,j}-1} }{\left( u^{\prime}_{\alpha,j} \right)^{n_{\alpha,j}} } \mathbb{I}(v_j \leq u_{\alpha,j}^{\prime}) \mathbb{I}(v_j \leq u_{\alpha,j}) d \mathbf{v} \\
=&
\int_{\mathbb{R}_{+}^{J}} \prod_{j=1}^J
\frac{ n_{\alpha,j} v_j^{n_{\alpha,j}-1} }{\left( u^{\prime}_{\alpha,j} \right)^{n_{\alpha,j}} } \mathbb{I}(v_j \leq u_{\alpha,j}^{\prime} \wedge u_{\alpha,j}) d \mathbf{v} \\
=&
\prod_{j=1}^J \int_{0}^{(u_{\alpha,j}^{\prime} \wedge u_{\alpha,j})}
\frac{ n_{\alpha,j} v_j^{n_{\alpha,j}-1} }{ \left( u^{\prime}_{\alpha,j} \right)^{n_{\alpha,j}} } d {v_j} \\
=&
\prod_{j=1}^J
\left\{ \frac{(u_{\alpha,j}^{\prime} \wedge u_{\alpha,j})}{u_{\alpha,j}^{\prime}} \right\}^{n_{\alpha,j}} \\
\geq &
\prod_{j=1}^J
\left( \frac{ u_{\alpha,j}}{ \sup u_{\alpha,j} } \right)^{n_{\alpha,j}} \\
= &
\prod_{j=1}^J
\left( \frac{ u_{\alpha,j}}{u_{\alpha,j}^{+}} \right)^{n_{\alpha,j}}
\end{array}
$$
where $\wedge$ denotes minimum. We may also use the $D_t \mathbf{z}^{\top}_{\alpha}(t_i) \mathbf{u}_{\alpha}$ notation in our proof instead. For our Mtd, we have 
$$
\begin{array}{rl}
k\left( \boldsymbol{\eta} \mid \boldsymbol{\eta}^{\prime}\right)
\geq & (2 \pi)^{-\frac{P+M}{2}}\left|\mathbf{A}\left(\tau_0\right)\right|^{1 / 2} R_1 \left(\frac{b_0}{b_0+\mathbf{u}_{\mathcal{B}}^\top \mathbf{u}_{\mathcal{B}} / 2}\right)^{a_0+M / 2} \mathbb{I}( \boldsymbol{\eta} \in \mathcal{C}_0 ) 
\left\{  \prod_{j=1}^J
\left( \frac{ u_{\alpha,j}}{u_{\alpha,j}^{+}} \right)^{n_{\alpha,j}} \right\} \\
& \times \exp \left( -\frac{1}{2} \left\{ \boldsymbol{\eta} - \mathbf{S}^{-1} \boldsymbol{\mu}_\Lambda \right\}^{\top} \mathbf{S} \left\{ \boldsymbol{\eta} - \mathbf{S}^{-1} \boldsymbol{\mu}_\Lambda \right\} \right) \\
= & \delta h(\boldsymbol{\eta})
\end{array}
$$
with 
$$
\begin{array}{rl}
\delta = & 
{c_3(\mathbf{M}, \mathbf{y})} 
(2 \pi)^{-\frac{P+M}{2}} \left|\mathbf{A}\left(\tau_0\right)\right|^{1 / 2} R_1 \\
h(\boldsymbol{\eta}) = & 
\frac{1}{c_3(\mathbf{M}, \mathbf{y})} 
\left(\frac{b_0}{b_0+\mathbf{u}_{\mathcal{B}}^\top \mathbf{u}_{\mathcal{B}} / 2}\right)^{a_0+M / 2} \mathbb{I}( \boldsymbol{\eta} \in \mathcal{C}_0 ) 
\left\{ \prod_{j=1}^J
\left( \frac{ u_{\alpha,j}}{u_{\alpha,j}^{+}} \right)^{n_{\alpha,j}} \right\} \\
& \times \exp \left( -\frac{1}{2} \left\{ \boldsymbol{\eta} - \mathbf{S}^{-1} \boldsymbol{\mu}_\Lambda \right\}^{\top} \mathbf{S} \left\{ \boldsymbol{\eta} - \mathbf{S}^{-1} \boldsymbol{\mu}_\Lambda \right\} \right) \\
c_3(\mathbf{M}, \mathbf{y}) = & 
\int_{\mathbb{R}^{P+M}} \Bigg[ \left(\frac{b_0}{b_0+\mathbf{u}_{\mathcal{B}}^\top \mathbf{u}_{\mathcal{B}} / 2}\right)^{a_0+M / 2} \mathbb{I}( \boldsymbol{\eta} \in \mathcal{C}_0 ) 
\left\{ \prod_{j=1}^J
\left( \frac{ u_{\alpha,j}}{u_{\alpha,j}^{+}} \right)^{n_{\alpha,j}} \right\} \\
& \times \exp \left( -\frac{1}{2} \left\{ \boldsymbol{\eta} - \mathbf{S}^{-1} \boldsymbol{\mu}_\Lambda \right\}^{\top} \mathbf{S} \left\{ \boldsymbol{\eta} - \mathbf{S}^{-1} \boldsymbol{\mu}_\Lambda \right\} \right) \Bigg] d \boldsymbol{\eta} \\
\leq & 
\int_{\mathbb{R}^{P+M}} 
\exp \left( -\frac{1}{2} \left\{ \boldsymbol{\eta} - \mathbf{S}^{-1} \boldsymbol{\mu}_\Lambda \right\}^{\top} \mathbf{S} \left\{ \boldsymbol{\eta} - \mathbf{S}^{-1} \boldsymbol{\mu}_\Lambda \right\} \right) d \boldsymbol{\eta} \\
= & (2 \pi)^{\frac{P+M}{2}} | \mathbf{S} |^{-1/2} .
\end{array}
$$
Note that $| \mathbf{S} |^{-1/2} |\mathbf{A}\left(\tau_0\right)|^{1/2} \leq 1$ and $\delta < 1$.

\end{proof}

\begin{proof}Proof of Theorem 2.
First, we define normalizing constant $c(\mathbf{y})$ and
$$
\begin{array}{c}
c(\mathbf{y})
=
\int_{\mathbb{R}_+} \int_{\mathbb{R}^{P+M}} \pi \left(\tau_{\mathcal{B}} \mid a_{0}, b_{0}, \tau_{0}\right) \pi( \boldsymbol{\beta} \mid \boldsymbol{\mu}_0, \boldsymbol{\Sigma}_0, \mathcal{C}_0 ) \pi( \mathbf{u}_{\mathcal{B}} \mid \mathbf{0}, \tau_{\mathcal{B}}^{-1} \mathbf{I}_{M} ) L(\boldsymbol{\eta} \mid \mathbf{y}, \mathbf{M} ) d \boldsymbol{\eta} d \tau_{\mathcal{B}} \\
\pi(\boldsymbol{\eta}, \tau_{\mathcal{B}}, \boldsymbol{\omega} \mid \mathbf{y})
=
c( \mathbf{y} )^{-1}{ \pi( \boldsymbol{\omega} \mid \boldsymbol{\eta}) L(\boldsymbol{\eta} \mid \mathbf{y}, \mathbf{M} ) \pi( \boldsymbol{\beta} \mid \boldsymbol{\mu}_0, \boldsymbol{\Sigma}_0, \mathcal{C}_0 ) \pi( \mathbf{u}_{\mathcal{B}} \mid \mathbf{0}, \tau_{\mathcal{B}}^{-1} \mathbf{I}_{M} ) \pi \left(\tau_{\mathcal{B}} \mid a_{0}, b_{0}, \tau_{0}\right)}\\
\pi(\boldsymbol{\eta} \mid \mathbf{y})= \int_{\mathbb{R}_+} \int_{\mathbb{R}_+^N} \pi(\boldsymbol{\eta}, \tau_{\mathcal{B}}, \boldsymbol{\omega} \mid \mathbf{y}) d \boldsymbol{\omega} d \tau_{\mathcal{B}}.
\end{array}
$$
We can integrate and bound the the slice variable at $\mathbf{u}_\alpha^+$. We can rewrite with kernels and bound the indicator function $\mathbb{I}( \boldsymbol{\eta} \in \mathcal{C}_0 ) \leq 1$, 
$$
\begin{array}{rl}
\pi(\boldsymbol{\eta}, \tau_{\mathcal{B}}, \boldsymbol{\omega} \mid \mathbf{y}) \leq & c(\mathbf{y})^{-1} \prod_{i=1}^N \left\{ D_t \mathbf{z}^{\top}_{\alpha}(t_i) \mathbf{u}^+_{\alpha} \right\}^{y_i} \\
& \times \prod_{i=1}^N e^{\kappa_i \psi_{i}} e^{-\omega_{i} \psi_{i}^2 / 2} \pi \left(\omega_{i} \mid y_i + \epsilon, 0\right) \\
& \times \exp \left\{ - \frac{1}{2} ( \boldsymbol{\beta} - \boldsymbol{\mu}_0 )^\top \boldsymbol{\Sigma}_0^{-1} ( \boldsymbol{\beta} - \boldsymbol{\mu}_0 )  \right\} \\
& \times \exp \left( - \frac{\tau_{\mathcal{B}}}{2} \mathbf{u}_\alpha^\top \mathbf{u}_\alpha  \right) \tau_{\mathcal{B}}^{M/2} \tau_{\mathcal{B}}^{a_{0}-1} \exp \left(-b_{0} \tau_{\mathcal{B}} \right) \mathbb{I}\left(\tau_{\mathcal{B}} \geq \tau_{0}\right) \\
= & c(\mathbf{y})^{-1} \prod_{i=1}^N \left\{ D_t \mathbf{z}^{\top}_{\alpha}(t_i) \mathbf{u}^+_{\alpha} \right\}^{y_i} \\
& \times \prod_{i=1}^N 
e^{ \kappa_i \eta_\epsilon } e^{\kappa_i \varphi_{i}} e^{-\omega_{i} \psi_{i}^2 / 2}
\pi \left(\omega_{i} \mid y_i + \epsilon, 0\right) \\
& \times \exp \left\{ - \frac{1}{2} ( \boldsymbol{\eta} - \mathbf{b} )^\top \mathbf{A}(\tau_{\mathcal{B}}) ( \boldsymbol{\eta} - \mathbf{b} )  \right\} \\
&\times \tau_{\mathcal{B}}^{M/2 + a_{0}-1} \exp \left(-b_{0} \tau_{\mathcal{B}} \right) \mathbb{I}\left(\tau_{\mathcal{B}} \geq \tau_{0}\right) \\
\leq & 
c(\mathbf{y})^{-1} \prod_{i=1}^N \left\{ D_t \mathbf{z}^{\top}_{\alpha}(t_i) \mathbf{u}^+_{\alpha} \right\}^{y_i} \\
& \times \prod_{i=1}^N 
e^{ \kappa_i \eta_\epsilon } e^{\kappa_i \varphi_{i}} 
\pi \left(\omega_{i} \mid y_i + \epsilon, 0\right) \\
& \times \exp \left\{ - \frac{1}{2} ( \boldsymbol{\eta} - \mathbf{b} )^\top \mathbf{A}(\tau_{\mathcal{B}}) ( \boldsymbol{\eta} - \mathbf{b} )  \right\} \\
&\times \tau_{\mathcal{B}}^{M/2 + a_{0}-1} \exp \left(-b_{0} \tau_{\mathcal{B}} \right) \mathbb{I}\left(\tau_{\mathcal{B}} \geq \tau_{0}\right) \\
= &
c(\mathbf{y})^{-1} \prod_{i=1}^N \left\{ D_t \mathbf{z}^{\top}_{\alpha}(t_i) \mathbf{u}^+_{\alpha} \right\}^{y_i} \\
& \times \prod_{i=1}^N 
e^{ \kappa_i \eta_\epsilon } 
\pi \left(\omega_{i} \mid y_i + \epsilon, 0\right) \\
& \times \exp \left( \boldsymbol{\kappa}^\top \mathbf{M} \boldsymbol{\eta} \right) \exp \left\{ - \frac{1}{2} ( \boldsymbol{\eta} - \mathbf{b} )^\top \mathbf{A}(\tau_{\mathcal{B}}) ( \boldsymbol{\eta} - \mathbf{b} )  \right\} \\
&\times \tau_{\mathcal{B}}^{M/2 + a_{0}-1} \exp \left(-b_{0} \tau_{\mathcal{B}} \right) \mathbb{I}\left(\tau_{\mathcal{B}} \geq \tau_{0}\right) .
\end{array}
$$
Next, 
$$
\begin{array}{rl}
\pi(\boldsymbol{\eta}, \tau_{\mathcal{B}}, \boldsymbol{\omega} \mid \mathbf{y}) 
\leq & 
c(\mathbf{y})^{-1} \prod_{i=1}^N \left\{ D_t \mathbf{z}^{\top}_{\alpha}(t_i) \mathbf{u}^+_{\alpha} \right\}^{y_i} \\
& \times \prod_{i=1}^N 
e^{ \kappa_i \eta_\epsilon } 
\pi \left(\omega_{i} \mid y_i + \epsilon, 0\right) \\
& \times \exp \left( \boldsymbol{\kappa}^\top \mathbf{M} \boldsymbol{\eta} \right) \frac{ \left| \mathbf{A}(\tau_{\mathcal{B}}) \right|^{1/2} }{ \left| \mathbf{A}(\tau_{\mathcal{B}}) \right|^{1/2} } \exp \left\{ - \frac{1}{2} ( \boldsymbol{\eta} - \mathbf{b} )^\top \mathbf{A}(\tau_{\mathcal{B}}) ( \boldsymbol{\eta} - \mathbf{b} )  \right\} \\
&\times \tau_{\mathcal{B}}^{M/2 + a_{0}-1} \exp \left(-b_{0} \tau_{\mathcal{B}} \right) \mathbb{I}\left(\tau_{\mathcal{B}} \geq \tau_{0}\right) \\
\leq &
c(\mathbf{y})^{-1} \left( 2 \pi \right)^{(P+M)/2} \prod_{i=1}^N \left\{ D_t \mathbf{z}^{\top}_{\alpha}(t_i) \mathbf{u}^+_{\alpha} \right\}^{y_i} \\
& \times \prod_{i=1}^N 
e^{ \kappa_i \eta_\epsilon } 
\pi \left(\omega_{i} \mid y_i + \epsilon, 0\right) \\
& \times \exp \left( \boldsymbol{\kappa}^\top \mathbf{M} \boldsymbol{\eta} \right) { \left| \mathbf{A}(\tau_{0}) \right|^{-1/2} } \pi(\boldsymbol{\eta} | \mathbf{b}, \mathbf{A}(\tau_{\mathcal{B}})^{-1} ) \\
&\times \tau_{\mathcal{B}}^{M/2 + a_{0}-1} \exp \left(-b_{0} \tau_{\mathcal{B}} \right) \mathbb{I}\left(\tau_{\mathcal{B}} \geq \tau_{0}\right) .
\end{array}
$$
Note that $|\mathbf{A}(\tau_{\mathcal{B}})|^{-1/2} \leq |\mathbf{A}(\tau_{0})|^{-1/2}$.
We set $\mathbf{s}^\top = \boldsymbol{\kappa}^\top \mathbf{M} + \mathbf{t}^\top$ and apply the MGF to normal distribution $\pi \left( \boldsymbol{\eta} | \mathbf{b}, \mathbf{A}(\tau_{\mathcal{B}})^{-1} \right)$ to get
$$
\begin{array}{rl}
\int_{\mathbb{R}^{P+M}} \exp( \mathbf{t}^\top \boldsymbol{\eta} ) \exp( \boldsymbol{\kappa}^\top \mathbf{M} \boldsymbol{\eta} ) \pi \left( \boldsymbol{\eta} | \mathbf{b}, \mathbf{A}(\tau_{\mathcal{B}})^{-1} \right) d \boldsymbol{\eta}
=& 
\exp \left[ \mathbf{s}^\top \mathbf{b} +\frac{1}{2} \mathbf{s}^\top \mathbf{A}(\tau_{\mathcal{B}})^{-1} \mathbf{s} \right] \\
\leq &
\exp \left[ \mathbf{s}^\top \mathbf{b} +\frac{1}{2} \mathbf{s}^\top \mathbf{A}(\tau_{0})^{-1} \mathbf{s} \right].
\end{array}
$$
All that is left is to integrate the gamma kernel and integrate Polya-gamma kernels. An upper bound is achieved and we have
$$
\begin{array}{c}
\int_{\mathbb{R}^{P+M}} e^{\boldsymbol{\eta}^{\top} \mathbf{t}} \pi(\boldsymbol{\eta} \mid \mathbf{y}) d \boldsymbol{\eta} = \int_{\mathbb{R}_+} \int_{\mathbb{R}_+^{N}} \int_{\mathbb{R}^{P+M}} e^{\boldsymbol{\eta}^{\top} \mathbf{t}} \pi(\boldsymbol{\eta}, \tau_{\mathcal{B}}, \boldsymbol{\omega} \mid \mathbf{y}) d\boldsymbol{\eta} d\boldsymbol{\omega} d\tau_{\mathcal{B}} <\infty.
\end{array}
$$
\end{proof}

\end{document}